\documentclass{ptephy_v1}

\preprintnumber{RIKEN-QHP-417} 

\usepackage{amsmath,amssymb,bm,yfonts,enumerate,color,natbib}
\usepackage{graphics}

\newcommand{\tr}{\mathop{\mathrm{tr}}}
\newcommand{\rank}{\mathop{\mathrm{rank}}}
\newcommand{\brokenPotential}{h}
\newcommand{\ri}{\mathrm{i}}
\newcommand{\average}[1]{\langle#1\rangle}

\begin{document}

\title{Spontaneous symmetry breaking and Nambu--Goldstone modes in open classical and quantum systems}

\author[1]{Yoshimasa Hidaka}
\affil{Nishina Center, RIKEN, Wako 351-0198, Japan,\\
RIKEN iTHEMS, RIKEN, Wako 351-0198, Japan \email{hidaka@riken.jp}}

\author[2]{Yuki Minami}
\affil{Department of Physics, Zhejiang University, Hangzhou 310027, China \email{yminami@zju.edu.cn} }

\begin{abstract}
We discuss spontaneous symmetry breaking of open classical and quantum systems.
When a continuous symmetry is spontaneously broken in an open system, a gapless excitation mode appears corresponding to the Nambu--Goldstone mode.
Unlike isolated systems, the gapless mode is not always a propagation mode, but it is a diffusion one.
Using the Ward--Takahashi identity and the effective action formalism, we establish the Nambu--Goldstone theorem in open systems, and
derive the low-energy coefficients that determine the dispersion relation of Nambu--Goldstone modes.
Using these coefficients, we classify the Nambu--Goldstone modes into four types: type-A propagation, type-A diffusion, type-B propagation, and type-B diffusion modes.
\end{abstract}

\maketitle

\section{Introduction}
\label{sec:intro}
Spontaneous symmetry breaking is one of the most important notions in modern physics.
When a continuous symmetry is spontaneously broken, a gapless mode appears called the Nambu--Goldstone (NG) mode~\cite{nambu1961dynamical, goldstone1961field, goldstone1962broken},
which governs the low-energy behavior of the system.  For example, a phonon in a crystal, which is the NG mode associated with translational symmetry breaking,  determines the behavior of the specific heat at low temperatures, which is nothing but the Debye $T^3$ law.

The nature of the NG modes such as the dispersion relations and the number of modes is determined by symmetry and its breaking pattern.
The relation between them was first shown for relativistic systems~\cite{nambu1961dynamical, goldstone1961field, goldstone1962broken}, where the number of NG modes is equal to the number of broken symmetries or generators.
It was extended to isolated systems without Lorentz symmetry~\cite{Watanabe:2012hr, Hidaka:2012ym, Watanabe:2014fva,Hayata:2014yga,Takahashi:2014vua}, where the number of NG modes does not coincide with the number of broken symmetries~\cite{Nielsen:1975hm, Miransky:2001tw, Schafer:2001bq}.
A typical example is a magnon in a ferromagnet, which is the NG mode associated with spontaneous breaking of the spin symmetry $O(3)$ to $O(2)$. The number of broken symmetries, $\dim(O(3)/O(2))$, is equal to two, but  only one magnon with  quadratic dispersion appears.
In general, when a global internal symmetry $G$ is spontaneously broken into its subgroup $H$, the number of NG modes is expressed as~\cite{Watanabe:2012hr, Hidaka:2012ym, Watanabe:2014fva,Hayata:2014yga,Takahashi:2014vua}
\begin{equation}
\begin{split}
N_\text{NG}=N_\text{BS} - \frac{1}{2}\rank\rho,
\label{eq:NGRelation}
\end{split}
\end{equation}
where $N_\text{BS}=\mathrm{dim}(G/H)$ is the number of broken symmetries, and $\rho^{\beta\alpha}:=-\langle[\ri \hat{Q}^\beta,\hat{j}^{\alpha 0}(x)]\rangle$ is the Watanabe--Brauner matrix with the Noether charge $Q^{\beta}$ and the charge density $j^{\alpha0}(x)$ of  $G$~\cite{Watanabe:2011ec}. 
The NG modes are classified into two types: type-A and type-B modes. The type-B mode is characterized by nonvanishing $\rho^{\beta\alpha}$ which implies that the two broken charge densities $j^{\beta0}(x)$ and $j^{\alpha0}(x)$ are not canonically independent~\cite{Nambu:2004}. $(\rank\rho)/2$ counts the number of canonical pairs of broken generators, and every pair of broken charges constitutes one NG mode. Therefore, the number of type-B modes, $N_\text{B}$, is equal to $(\rank\rho)/2$. 
The rest of the degrees of freedom, $N_\text{A} = N_\text{BS}-\rank\rho$, become the type-A modes, which have the same property as NG modes in Lorentz-invariant systems. The sum of the number of type-A and type-B modes leads to Eq.~\eqref{eq:NGRelation}. Both type-A and type-B NG modes are  propagation modes with typically linear and quadratic dispersions, respectively.

Similar to isolated systems, spontaneous symmetry breaking occurs in open systems. A well-known example is 
a diffusion mode in the synchronization transition of coupled oscillators, which describe chemical and biological oscillation phenomena~\cite{kuramoto2012chemical, Acebron:2005zz}.
In the synchronization transition, $U(1)$ phase symmetry is spontaneously broken, and the diffusion mode appears as the NG mode, which has different dispersion from that in isolated systems.
Another example is ultracold atoms in an optical cavity~\cite{baumann2010dicke}. 
In the system, a laser and its coupling to radiation fields give rise to spontaneous emission and dissipation, where the internal energy does not conserve.
Bose--Einstein condensation and the symmetry breaking can occur even in such a case, and they have been observed~\cite{baumann2010dicke}.
Furthermore,  the synchronization transition and the diffusive NG mode of  ultracold atoms in the driven-dissipative setup are discussed in Refs.~\cite{PhysRevB.74.245316,PhysRevLett.99.140402,PhysRevA.76.043807,PhysRevLett.96.230602,Sieberer:2015svu,Minami:2018oxl}.
The diffusive NG modes are characteristic of open systems. 

Compared with isolated systems,  the relation between the NG modes, the broken symmetries, and the dispersion relations has not been established in open systems. 
A crucial difference between isolated and open systems is the lack of ordinary conserved quantities such as the energy, momentum, and particle number, because of the interaction with the environment. Thus, we cannot naively apply the argument for isolated systems to that for open systems. 
In our previous work, we studied properties of the NG modes in open systems based on toy models~\cite{Minami:2018oxl},
in which we found two types of NG modes: diffusion and propagation modes, whose poles have $\omega = -\ri \gamma |\bm{k}|^{2}$ and $(\pm a-\ri b)|\bm{k}|^{2}$, respectively. Here, $\gamma$, $a$, and $b$ are constants.  In the model study, the nonvanishing Watanabe--Brauner matrix for the open system leads to the propagation NG modes, where a similar relation to Eq.~\eqref{eq:NGRelation} is satisfied~\cite{Minami:2018oxl}.
However, it is found that the propagation modes split into two diffusion modes by changing a model parameter in the study of time-translation breaking~\cite{Hayata:2018qgt}, where Eq.~\eqref{eq:NGRelation} is not always satisfied. Therefore, we need a model-independent analysis to understand the nature of NG modes for open classical and quantum systems. This is the purpose of the present paper.

Open classical and quantum systems can be uniformly described by the path integral formulation, called the Martin--Siggia--Rose--Janssen--De Dominicis (MSRJD) formalism~\cite{MSR, J, D1, D2} for classical systems, and the Keldysh formalism for quantum systems~\cite{keldysh1965diagram}.
Both path integral formulations are written in two fields. In the language of classical theories, they are called classical and response fields.
In the Keldysh formalism, they correspond to the a linear combination of fields on the forward and backward paths.
These doubled fields play an important role in the symmetry of open systems. 

The notion of symmetry in open systems is slightly different from that in isolated systems~\cite{Sieberer:2015svu,Minami:2018oxl, Hayata:2018qgt}.
In isolated systems, when there is a continuous symmetry, there exists a physical Noether charge, which we call  $Q_{R}^{\alpha}$ in this paper. 
``Physical'' means that the Noether charge is an observable like the energy and momentum.
In contrast, in open systems, the Noether theorem does not necessarily lead to the physical conserved charge. Instead, another conserved charge, which we refer to as $Q_{A}^{\alpha}$, arises in the path integral formulation. The doubled charges, $Q_{R}^{\alpha}$ and $Q_{A}^{\alpha}$, relate to the fact that the path integral is written in doubled fields.
Although $Q_{A}^{\alpha}$ itself is not a physical conserved quantity, it plays the role of the symmetry generator.
By using $Q_{A}^{\alpha}$, we can define the spontaneous symmetry breaking for open systems and derive the Ward--Takahashi identities~\cite{PhysRev.78.182,Takahashi1957}.
As is the case in isolated systems, spontaneous symmetry breaking implies  the existence of gapless excitation modes. 
Our previous analysis based on the Ward--Takahashi identities~\cite{Minami:2018oxl} is limited in the zero-momentum limit. 
In this paper, we generalize it to analysis with finite momentum,
and derive the low-energy coefficients for the inverse of the Green functions in the NG mode channel.  Using these coefficients, we classify the NG modes, and discuss the relation between these modes and the broken generators.

The paper is organized as follows: 
In Sec.~\ref{sec:mainResult}, we show our main result, in which we classify the NG modes into four types and discuss their dispersion relations.
We also discuss how the NG modes can be observed in experiments.
In Sec.~\ref{sec:OCQS}, we review the path integral formulation in open classical and quantum systems.
In Sec.~\ref{sec:technique}, we discuss the concept of symmetry in open systems and provide two field-theoretical technique that we employ to show our main result.
Section~\ref{sec:SSB} shows the detailed derivation of our main result.
Section~\ref{sec:summary} is devoted to the summary and discussion.

Throughout this paper, we use the relativistic notation in $(d+1)$-dimensional spacetime with the Minkowski metric $\eta_{\mu\nu}=\mathrm{diag}(1,-1,-1,\cdots, -1)$,
although the system need not have the Lorentz symmetry. 
We employ the natural units, i.e., $c=\hbar=1$, where $c$ is the speed of light and $\hbar$ is the Planck constant over $2\pi$.

\section{Main result}\label{sec:mainResult}
In this section we summarize our main results first, because the derivation is technical and complicated.
We show the relationship between the inverse of the retarded Green function and low-energy coefficients that determine the dispersion relation of NG modes.
Using  the low-energy coefficients, we classify the NG modes into four types: type-A propagation, type-A diffusion, type-B propagation, and type-B diffusion modes. We also discuss how these modes can be observed in experiments.
\subsection{Green functions and low-energy coefficients}\label{sec:GreenFunctionsLow-energyCoefficients}
We consider a $(d+1)$-dimensional open system that has a continuous internal symmetry $G$ (the precise definition of symmetry in open systems is shown in Sec.~\ref{sec:symmetry}).
Suppose that the symmetry is spontaneously broken into its subgroup $H$ in a steady state.
We assume that the steady state is unique and stable against small perturbations.
The steady state may be not only the global thermal equilibrium but also a nonequilibrium steady state. We also assume that any spacetime symmetries of the open system such as 
the time and spatial translational symmetry are not spontaneously broken, which implies that the frequency and momentum are good quantum numbers.
We are interested in the behavior of the retarded Green functions that contain NG modes $[G_{\pi}(k)]_{\alpha\beta}$, where $\alpha$ and $\beta$ run from $1$ to the number of broken symmetries $N_\text{BS}$. As in the case of isolated systems, $N_\text{BS}$ is equal to $\dim(G/H)$.
 Our main result (generalization of  the Nambu--Goldstone theorem) is that the inverse of the retarded Green function can be expanded as
\begin{equation}
\begin{split}
[G_\pi^{-1}(k)]^{\beta\alpha} = C^{\beta\alpha}   -\ri C^{\beta\alpha;\mu} k_\mu +C^{\beta\alpha;\nu \mu}k_\nu k_\mu +\cdots,
\label{eq:invG}
\end{split}
\end{equation}
with coefficients:
\begin{align}
C^{\beta\alpha} &= 0, \label{eq:Cformula}\\
  C^{\beta\alpha;\mu}
&=
\average{\delta_{R}^{\beta} j_{A}^{\alpha\mu}(0)}
+ \ri\int d^{d+1}x\,\langle \bigl(\mathcal{Q}_{\pi}\brokenPotential_{R}^{\beta}(x)\bigr)\,\big(\mathcal{Q}_{\pi}j_{A}^{\alpha\mu}(0)\bigr)\rangle_\text{c}
, \label{eq:C00formula}\\
  C^{\beta\alpha;\nu \mu}
&=
\average{ \mathcal{S}^{\beta\alpha;\nu\mu}(0)}
-\ri\int d^{d+1}x\, \average{    \bigl(\mathcal{Q}_\pi j^{\beta\nu}_R(x)\bigr)\,\bigl( \mathcal{Q}_\pi j^{\alpha\mu}_A(0)\bigr) }_{\text{c}}\\
&\quad+\ri\int d^{d+1}x\,  x^{\nu}\langle{ \bigl(\mathcal{Q}_\pi\brokenPotential_{R}^{\beta}(x)\bigr)\,\bigl(\mathcal{Q}_\pi j_A^{\alpha\mu}(0)\bigr)}\rangle_\text{c}.
\label{eq:C10formula}
\end{align}
Here, $\mu$ and $\nu$ are spacetime indices, $k^{\mu}=(\omega,\bm{k})$ are the frequency and the wave vector, {and $d^{d+1}x = dt d\bm{x}$.
The subscript  c denotes the connected part of correlation functions.
$\mathcal{Q}_\pi$ is the projection operator that removes the contribution of NG modes from the operator
[the definition of $\mathcal{Q}_\pi(=1-\mathcal{P}_\pi)$ is given in Eq.~\eqref{eq:Ppi}].
The expectation values are evaluated in the path integral formulation, Eq.~\eqref{eq:PathIntegral}.
In this formulation, there are two types elementary of fields denoted as $\chi_R^{a}$ and $\chi_A^{a}$, which correspond to the classical and fluctuation fields, respectively.
The operators in Eqs.~\eqref{eq:C00formula} and \eqref{eq:C10formula} are related to the symmetry transformation of the action $S[\chi_{R}^{a},\chi_{A}^{a}]$.
To see this, consider an infinitesimal local transformation of fields under $G$:
$\chi_{i}^{a}(x)\to \chi_{i}^{a}(x)+\delta_{A}\chi_{i}^{a}(x)$ $(i=R, A)$ with $\delta_{A}\chi_{i}^{a}(x) := \epsilon_{\alpha}(x)\delta_{A}^{\alpha}\chi_{i}^{a}(x)$, where 
$\epsilon_{\alpha}(x)$ is the infinitesimal parameter depending on spacetime, and $\delta_{A}^{\alpha}\chi_{i}^{a}(x)$ is the infinitesimal transformation of $\chi_{i}^{a}(x)$. 
If $\chi_{i}^{a}(x)$ is a linear representation of $G$, it can be represented as $\delta_{A}^{\alpha}\chi_{i}^{a}(x)= \ri [T^{\alpha}]^{a}_{~b}\chi_{i}^{b}(x)$, where  $T^{\alpha}$ is the generator of $G$.
Since $G$ is the symmetry, the action is invariant under $\delta_{A}$ if $\epsilon_{\alpha}$ is constant.
For a spacetime-dependent $\epsilon_{\alpha}(x)$, the action transforms as
\begin{equation}
\begin{split}
\delta_{A} S &=- \int d^{d+1}x\,\partial_{\mu}{\epsilon}_{\alpha}(x) j_{A}^{\alpha\mu}(x),
\label{eq:deltaAS}
\end{split}
\end{equation}
where $j_{A}^{\alpha\mu}(x)$ is the Noether current.
We also introduce a local transformation $\delta_{R}$ such that $\delta_{R}\chi_{i}^{a}=\bar{\epsilon}_{\alpha}(x)\delta_{R}^{\alpha}\chi_{i}^{a}(x)$, with 
\begin{equation}
\begin{split}
\delta_{R}^{\alpha}\chi_{A}^{a}(x) &:= \delta_{A}^{\alpha}\chi_{R}^{a}(x),\\
\delta_{R}^{\alpha}\chi_{R}^{a}(x)&: = 
\begin{cases}
 \frac{1}{4}\delta_{A}^{\alpha}\chi_{A}^{a}(x) & \text{quantum system}\\
 0 &\text{classical system}
 \end{cases}.
 \end{split}
  \label{eq:transR}
\end{equation}
In isolated systems, these are also symmetry transformations; however, they are not in open systems~\cite{Minami:2018oxl,Sieberer:2015svu}.
Under this transformation, the action transforms as
\begin{equation}
\delta_{R} S = \int d^{d+1}x\, \bigl[\bar{\epsilon}_{\alpha}(x) \brokenPotential_{R}^{\alpha}(x) 
-\partial_{\mu}\bar{\epsilon}_{\alpha}(x) j_{R}^{\alpha\mu}(x) 
\bigr].
\label{eq:deltaRS3}
\end{equation}
Since $\delta_{R}$ transformation is not the symmetry of the action, $\brokenPotential_{R}^{\alpha}(x) $ exists.

$\delta_{R}^{\beta}j_{A}^{\alpha\mu}(x)$ and $\mathcal{S}^{\beta\alpha;\nu\mu}(x)$ in Eqs.~\eqref{eq:C00formula} and \eqref{eq:C10formula} are given through the infinitesimal local transformation of $j_{A}^{\alpha\mu}(x)$, which is
\begin{equation}
\delta_{R}j_{A}^{\alpha\mu}(x) = 
\bar{\epsilon}_{\beta}(x)\delta_{R}^{\beta}j_{A}^{\alpha\mu}(x)+
\partial_{\nu}\bar{\epsilon}_{\beta}(x) \mathcal{S}^{\beta\alpha;\nu\mu}(x)+\cdots.
\label{eq:deltaRJA}
\end{equation}
The low-energy coefficients in the inverse of retarded Green functions in Eqs.~\eqref{eq:Cformula}-\eqref{eq:C10formula} are expressed as one- or two-point functions of these operators.

The ordinary Nambu--Goldstone theorem shows $C^{\beta\alpha}=0$  in \eqref{eq:Cformula}, which is derived from the only symmetry breaking pattern, and thus it is independent of the details of the underlying theory. This claims that there is at least one zero mode when a continuum symmetry is spontaneously broken.
To determine the dispersion relation and the number of NG modes, we need additional data. If we impose Lorentz invariance on the system, we find $C^{\beta\alpha;\mu}=0$ and $C^{\beta\alpha;\nu\mu}=-\eta^{\nu\mu}g^{\beta\alpha}$, where $g^{\beta\alpha}$ is an $N_\text{BS}\times N_\text{BS}$ matrix with $\det g\neq0$.  In this case, $\det G_{\pi}^{-1}=0$ has $2N_\text{BS}$ solutions with $\omega=\pm|\bm{k}|$.  Each pair of solutions $\omega=\pm|\bm{k}|$ gives one mode. Therefore, the number of NG modes is equal to $N_\text{BS}$, whose dispersion is linear. This is the Nambu--Goldstone theorem in relativistic systems~\cite{goldstone1962broken,WeinbergText}. 
Equations~\eqref{eq:C00formula} and \eqref{eq:C10formula} give the data for general cases, not just for isolated systems without Lorentz invariance but also for open ones. In this sense, our formulae in Eqs.~\eqref{eq:invG}-\eqref{eq:C10formula} provide the generalization of the Nambu--Goldstone theorem.
In the next subsection, we discuss the classification of NG modes and their dispersion relations.

\subsection{Classification of NG modes}\label{sec:Classification}
It is interesting to clarify the relation between the broken symmetries, the NG modes, and their dispersion relations in open systems.
In isolated systems without Lorentz invariance, those relations have been made clear in Refs.~\cite{Watanabe:2012hr, Hidaka:2012ym, Watanabe:2014fva, Hayata:2014yga}. In this section, we generalize the relation obtained in isolated systems to that in open systems. 
The formula in isolated systems can be reproduced as a special case of our result.

Our formulae in Eqs.~\eqref{eq:invG}-\eqref{eq:C10formula} are quite general, so we need additional assumptions to classify NG modes. 
First, we assume that $\det(-\ri \omega \rho-\omega^{2}\bar{g})\neq0$ at an arbitrary small but nonzero $\omega$, where we define  $\rho^{\beta\alpha}:=C^{\beta\alpha;0}$ and $\bar{g}^{\beta\alpha}: = -C^{\beta\alpha;00}$.
This assumption means that the low-energy degrees of freedom are contained in, at least, up to quadratic order in $\omega$. 
We note that this is implicitly assumed in the analysis of NG modes in isolated systems~\cite{Watanabe:2012hr}.
Second, we assume that the action of the underlying theory satisfies the reality condition $\bigl(S[\chi^{a}_{R},\chi^{a}_{A} ] \bigr)^{*} =  - S[\chi^{a}_{R},-\chi^{a}_{A} ]$.    All the models discussed in Sec.~\ref{sec:OCQS} satisfy this condition.
The reality condition implies  that all the coefficients $C^{\beta\alpha;\mu\cdots}$ in Eq.~\eqref{eq:invG} are real.
This property leads to the relation that  if $\omega_{\bm{k}}$ is a solution of $\det G_{\pi}^{-1}(\omega,\bm{k})=0$, then $-\omega_{-\bm{k}}^{*}$ is also a solution.
The third assumption is that $G^{-1}_{\pi}(\omega,\bm{k})$ is invariant under $\bm{k}\to -\bm{k}$, which makes our analysis simpler. 
In this section, we concentrate on systems that satisfy this condition. One can generalize our analysis by relaxing the this assumption.
From these three assumptions, we can conclude that there are two types of modes: diffusion and propagation, and the poles of the retarded Green functions have the form:
\begin{equation}
\begin{split}
\omega =
\begin{cases}
  -\ri \gamma(\bm{k}) &\text{diffusion mode}\\
 \pm a(\bm{k}) -\ri b(\bm{k}) &\text{propagation mode}
\end{cases}.
\end{split}
\label{eq:diffusionPropagation}
\end{equation}
Here, $a$, $b$, and $\gamma$ are real and positive due to the stability of the system and they vanish at $\bm{k}=\bm{0}$ due to symmetry breaking.
We note that the above three assumptions do not exclude the possibility of the pole $\omega=0$ with finite momentum $\bm{k}$ or equivalently the pole with $a=0$, $b=0$, or $\gamma=0$. The pole can be excluded from the assumption that the translational symmetry is not broken discussed in Sec.~\ref{sec:GreenFunctionsLow-energyCoefficients}. 
If such a pole exists, there exists a time-independent local operator. By operating the operator to a given steady state, we can generate
another steady state with a finite momentum, which breaks the translational symmetry. This contradicts the assumption of broken translational invariance\footnote{The same argument is used in the analysis of NG modes in nonrelativistic systems~\cite{Nielsen:1975hm}. }.


Let us now focus on the relation between the number of modes and the coefficients $\rho^{\beta\alpha}$ and $\bar{g}^{\beta\alpha}$.
The number of zero modes can be evaluated from the number of solutions of $\det(-\ri \omega \rho-\omega^{2}\bar{g})=\omega^{N_\text{BS}}\det(-\ri \rho-\omega\bar{g}) =0$ with $\omega=0$, which is equal to $N_\text{BS}+(N_\text{BS}-\rank \rho)=2N_\text{BS}-\rank\rho$.
As in the case of isolated systems~\cite{Watanabe:2012hr}, we characterize the type-B modes by linearly independent  row vectors in $\rho$.
$\rho$ generally has both symmetry and antisymmetric parts. This is different from isolated systems, where $\rho$ is an antisymmetry matrix.
The symmetric part plays the role of dissipation. The number of type-B degrees of freedom is equal to $\rank\rho$.
This characterization is different from our previous work~\cite{Minami:2018oxl}, in which we proposed that the type-B mode was characterized by the antisymmetry part of $\rho$, and conjectured that the existence of the antisymmetric part implied the existence of a propagation mode. However, this conjecture is not always true~\cite{Hayata:2018qgt}; it depends on the details of the underlying theory, as will be seen below. Therefore, we change the definition of type-B modes\footnote{The type-A NG mode with $\omega= -\ri |\bm{k}|^{2}$ in our previous work~\cite{Minami:2018oxl} corresponds to the type-B diffusion mode in the new classification of the this paper.}.

Since propagation poles are always paired with the positive and negative real parts, we count the pair of poles as two. On the other hand, the pole of a diffusion mode is counted as one. This observation gives the relation between $\rank \rho$ and the number of diffusion and propagation modes as
\begin{equation}
\begin{split}
\rank \rho &= N_{\text{B-diffusion}} +2N_{\text{B-prop}}.
\label{eq:Type-B}
\end{split}
\end{equation}
As mentioned above, the number of diffusion modes depends on the details of the theory. 
To see this, let us consider a simple model with
\begin{equation}
\begin{split}
G^{-1}_{\pi}=
\begin{pmatrix}
-\ri\kappa_{1} \omega +\bm{k}^{2} & -\ri \omega\\
\ri \omega & -\ri \kappa_{2}\omega +\bm{k}^{2}
\end{pmatrix},
\qquad 
\rho=
\begin{pmatrix}
\kappa_{1} & 1\\
-1 & \kappa_{2}
\end{pmatrix},
\end{split}
\end{equation}
and $\bar{g}^{\beta\alpha}=0$.  Here, $\kappa_{1}$ and $\kappa_{2}$ are parameters. We choose these to be positive.
Since $\det \rho =\kappa_{1}\kappa_{2}+1\neq0$, $\rank \rho=2$, i.e., there are two solutions in $\det G^{-1}_{\pi}=0$,
which are given as
\begin{equation}
\begin{split}
\omega = \frac{-\ri(\kappa_{1}+\kappa_{2})\pm\sqrt{4-(\kappa_{1}-\kappa_{2})^{2}}}{2(1+\kappa_{1}\kappa_{2})}|\bm{k}|^{2}.
\end{split}
\end{equation}
If $4>(\kappa_{1}-\kappa_{2})^{2}$, one propagation mode appears.
On the other hand, if $4<(\kappa_{1}-\kappa_{2})^{2}$, two diffusion modes appear. In both cases, Eq.~\eqref{eq:Type-B} is, of course, satisfied;
however, the diffusion or propagation depends on the parameters.

The remaining of gapless degrees of freedom describe  type-A modes, whose number is $(N_\text{BS}-\rank \rho)$. Since they have $\omega^{2}$ terms in $G_{\pi}^{-1}$, we count each degree of freedom as two, and we find that the following relation is satisfied: 
\begin{equation}
\begin{split}
2(N_\text{BS}-\rank \rho) = N_{\text{A-diffusion}} +2N_{\text{A-prop}}.
\label{eq:Type-A}
\end{split}
\end{equation}
One might think the existence of the type-A diffusion mode is unnatural. However, we cannot exclude this possibility at this stage. For example, let us  consider $G_{\pi}^{-1}=-\omega^{2}-2\ri \zeta |\bm{k}|^{2}\omega+|\bm{k}|^{4}$. 
The solutions of $G_{\pi}^{-1}=0$ are $\omega =  -\ri \zeta|\bm{k}|^{2}\pm \sqrt{1-\zeta^{2}}|\bm{k}|^{2}$, which correspond to  the type-A modes because $\rho=0$.
If $\zeta<1$,  there is one propagation mode. On the other hand, if $\zeta>1$, there are  two diffusion modes.  These are nothing but the type-A diffusion modes in our classification. The possibility of type-A diffusion modes is excluded by assuming an additional condition, as will be seen later.

Combining Eqs.~\eqref{eq:Type-B} and \eqref{eq:Type-A}, we find the general relation:
\begin{equation}
\begin{split}
N_\text{BS} = N_{\text{A-prop}}+2N_{\text{B-prop}}+\frac{1}{2}N_{\text{A-diffusion}}+N_{\text{B-diffusion}}.
\label{eq:relation}
\end{split}
\end{equation}
For isolated systems, where $\brokenPotential_{R}^{\beta}$ vanishes,  the transformation generated by $\delta_{R}^{\beta}$ is upgraded to symmetry whose Noether current is $j^{\beta \mu}_{R}$.
In this case, the symmetric part of $\rho^{\beta\alpha}$ vanishes, and $\rho^{\beta\alpha}$ turns into the Watanabe--Brauner matrix: $\rho^{\beta\alpha}= \langle\delta_{R}^{\beta}j_{A}^{\alpha0}\rangle= -\langle[\ri \hat{Q}^{\beta}_{R},\hat{j}_{A}^{\alpha0}] \rangle$~\cite{Watanabe:2011ec}. In isolated systems, there are no diffusion modes, i.e., $N_{A\text{-diffusion}}=N_{B\text{-diffusion}}=0$, so the relation reduces to the standard one:
\begin{equation}
N_{\text{A-prop}}=N_\text{BS}-\rank\rho,\qquad
N_{\text{B-prop}}=\frac{1}{2}\rank\rho.
\end{equation}
Furthermore, let us consider that the system is isotropic. In this case, $C^{\alpha\beta;ij}$ can be expressed as $g^{\alpha\beta}\delta^{ij}$.
We assume that $\det g\neq0$, which simplifies the dispersion relations:
\begin{equation}
\begin{split}
\omega =
\begin{cases}
 \pm a_{B} |\bm{k}|^{2} -\ri b_{B}|\bm{k}|^{2} &\text{Type-B propagation mode}\\
  -\ri \gamma_{B} |\bm{k}|^{2}&\text{Type-B diffusion mode}\\
   \pm a_{A} |\bm{k}|-\ri b_{A}|\bm{k}|^{2}&\text{Type-A propagation mode}
\end{cases}.
\label{eq:classification}
\end{split}
\end{equation}
In this case, there is no type-A diffusion mode (see Appendix~\ref{sec:NoTypeADiffusion} for the proof).

In addition to gapless solutions, gapped solutions in $\det G_{\pi}^{-1}(\omega,\bm{k})=0$ may exist.
As is in the case of the gapless modes in Eq.~\eqref{eq:diffusionPropagation}, the gapped solutions can be classified into  damping and gapped modes, with 
\begin{equation}
\begin{split}\label{eq:gappedMode}
\omega =
\begin{cases}
  -\ri \gamma_{G} &\text{damping mode}\\
 \pm a_{G}-\ri b_{G}&\text{gapped mode}
\end{cases},
\end{split}
\end{equation}
at $\bm{k}=\bm{0}$.
The total number of solutions of $\det G_{\pi}^{-1}(\omega,\bm{0})=0$ is equal to the degree of the polynomial $\det G_{\pi}^{-1}(\omega,\bm{0})$.
Since $\det G_{\pi}^{-1}(\omega,\bm{0})=\det(-\ri \omega \rho-\omega^{2}\bar{g})=\omega^{N_\text{BS}}\det(-\ri\rho-\omega\bar{g})$, 
the degree is  equal to $(N_\text{BS}+\rank \bar{g})$.
There are $2N_\text{BS}-\rank\rho$ solutions with $\omega=0$, so that the remaining  $(\rank \bar{g}+N_\text{BS}) - (2N_\text{BS}-\rank\rho)=\rank \bar{g}+\rank \rho -N_\text{BS}$ solutions correspond to gapped ones.  Therefore, we find the relation for the gapped modes to be
\begin{equation}
\begin{split}
\rank \bar{g}+\rank \rho -N_\text{BS} =  2N_\text{gapped}+ N_\text{damping}.
\end{split}
\end{equation}
If $\gamma_{G}$ or $|a_{G}-\ri b_{G}|$ is much smaller than the typical energy scale, these gapped modes become the low-energy degrees of freedom.
In isolated systems, these kinds of gapped modes are known as gapped partners~\cite{Volkov:1972,Kapustin:2012cr,Hidaka:2012ym,Nicolis:2013sga,Gongyo:2014sra,Hayata:2014yga,Beekman:2014cba}. 
An example is a Ferrimagnet, in which  ferro- and antiferro- order parameters  coexist. 
The Ferrimagnet has one magnon with quadratic dispersion, which is the type-B mode, and one gapped mode.

\subsection{Spectral function and experimental detection}

The dispersion relation of NG modes can be experimentally observed by inelastic scattering processes~\cite{PhysRevLett.101.250403,PhysRevLett.88.120407,hoinka2017goldstone}. For example, in atomic Fermi superfluids, the NG mode is observed in the spectra with focused Bragg scattering~\cite{hoinka2017goldstone}. 
The differential cross section is proportional to the correlation function, which behaves as $2\mathrm{Im}\, G_{\pi}(\omega,\bm{k})/\omega=:S(\omega,\bm{k})$ at small $\omega$~\cite{Chaikin}\footnote{More precisely, the differential cross section is proportional to $G_{12}(\omega,\bm{k})$ in $1/2$ basis of Keldysh formalism.
At low $\omega$, $G_{12}(\omega,\bm{k})$ reduces to $2\mathrm{Im}\, G_{\pi}(\omega,\bm{k})/\omega$.}.
$S(\omega,\bm{k})$ will be multiplied by an additional factor depending on the processes. 
From the results in the previous subsection, the retarded Green functions in broken phases are written as
\begin{equation}
\begin{split}
G_\text{B-diffusion}(\omega,\bm{k}) &=  \frac{\ri\gamma_{B}}{\omega+\ri \gamma_{B}|\bm{k}|^{2}},\\
G_\text{B-prop}(\omega,\bm{k}) &=  \frac{-1}{(\omega-a_{B}|\bm{k}|^{2}+\ri b_{B}|\bm{k}|^{2})(\omega+a_{B}|\bm{k}|^{2}+\ri b_{B}|\bm{k}|^{2})}.
\end{split}
\end{equation}
The corresponding spectra $S_\text{B-diffusion}(\omega,\bm{k}):=2\mathrm{Im}\, G_\text{B-diffusion}(\omega,\bm{k}) /\omega$ and $S_\text{B-prop}(\omega,\bm{k}):=2\mathrm{Im}\, G_\text{B-prop}(\omega,\bm{k}) /\omega$ are 
\begin{align}
S_\text{B-diffusion}(\omega, \bm{k}) &=\frac{2\gamma_B}{\omega^2+\gamma_B^2|\bm{k}|^4}, \label{eq:spectraSAO}\\
S_\text{B-prop}(\omega, \bm{k}) &=\frac{4b_B|\bm{k}|^2}{\bigl((\omega-a_B|\bm{k}|^2)^2+b_B^2|\bm{k}|^4\bigr)\bigl((\omega+a_B|\bm{k}|^2)^2+b_B^2|\bm{k}|^4\bigr)},\label{eq:spectraSBO}
\end{align}
respectively.
As a comparison, we show the functional form of $S=\rho(\omega,\bm{k})/\omega$ for type-A and -B modes in isolated systems as
\begin{align}
S_A(\omega, \bm{k}) &:=\frac{4b_A|\bm{k}|^2}{\bigl((\omega-a_A|\bm{k}|)^2+b_A^2|\bm{k}|^4\bigr)\bigl((\omega+a_A|\bm{k}|)^2+b_A^2|\bm{k}|^4\bigr)},\\
S_B(\omega, \bm{k}) &=\frac{4b_B|\bm{k}|^4}{\bigl((\omega-a_B|\bm{k}|^2)^2+b_B^2|\bm{k}|^8\bigr)\bigl((\omega+a_B|\bm{k}|^2)^2+b_B^2|\bm{k}|^8\bigr)},
\end{align}
respectively.
\begin{figure}[htbp]
  \begin{center}
    \begin{tabular}{c}
      \begin{minipage}{0.5\hsize}
        \begin{center}
          \includegraphics[width=\hsize]{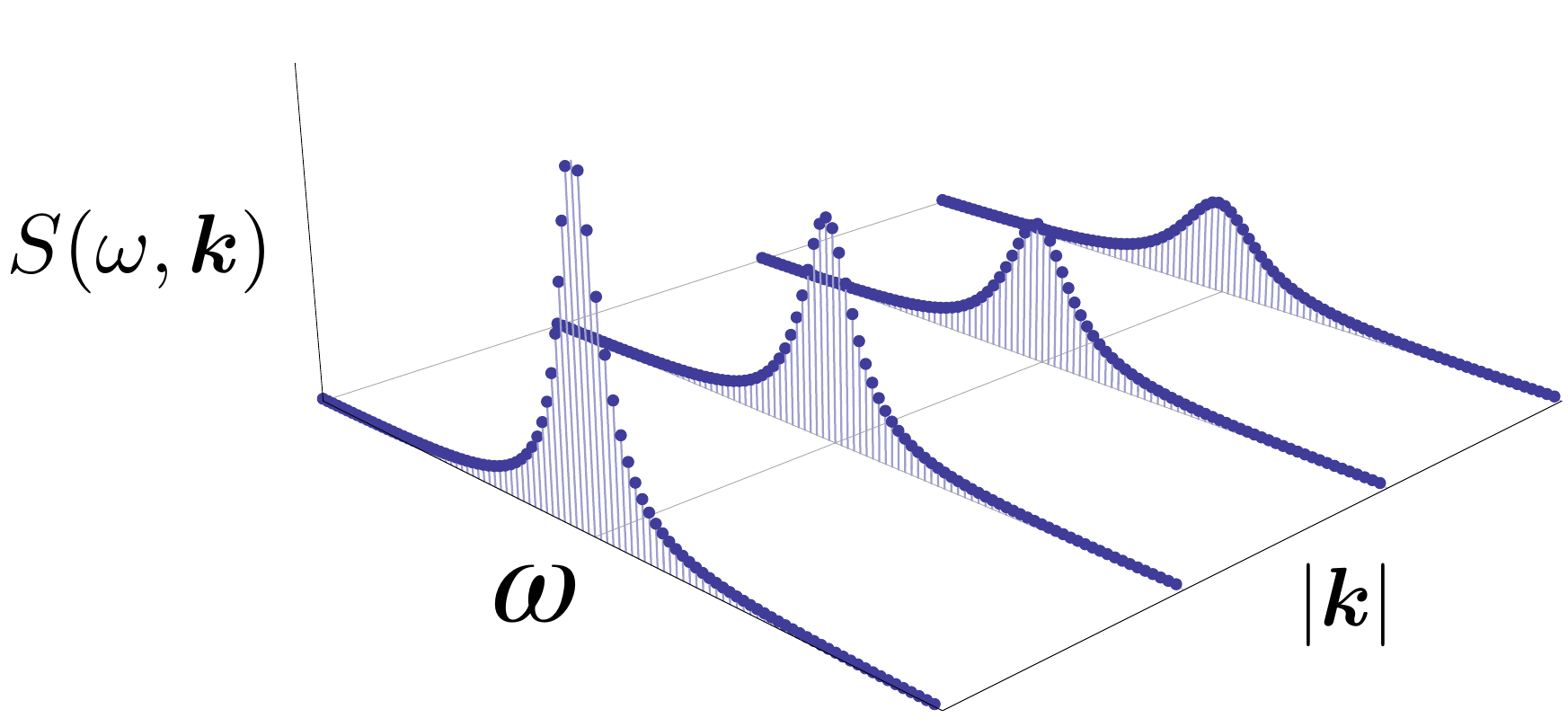}
        \end{center}
      \end{minipage}
      \begin{minipage}{0.5\hsize}
        \begin{center}
          \includegraphics[width=\hsize]{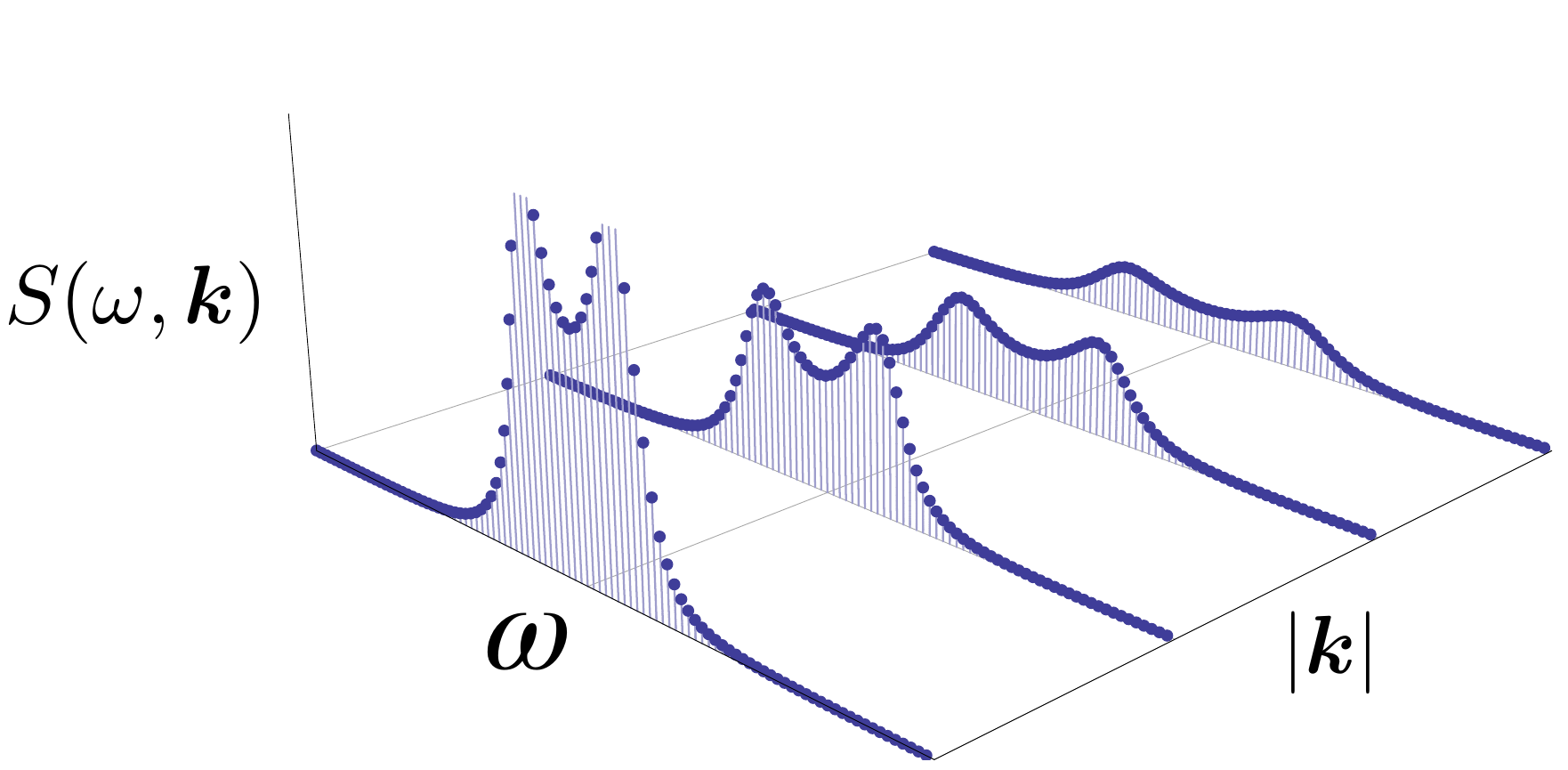}
        \end{center}
      \end{minipage}
    \end{tabular}
    \caption{The left panel shows the $\omega$-$|\bm{k}|$ dependence of $S(\omega,\bm{k})$ for a type-B diffusion mode, which  has a single peak. The right panel shows $S(\omega,\bm{k})$ for a type-B propagation mode, which has blunt pair peaks.  The parameters are chosen as $\gamma_{B}=1$, $a_{B}=1$, and $b_{B}=0.5$.}
    \label{fig:SpectralOpen}
  \end{center}
\end{figure}
\begin{figure}[htbp]
  \begin{center}
    \begin{tabular}{c}
      \begin{minipage}{0.5\hsize}
        \begin{center}
          \includegraphics[width=\hsize]{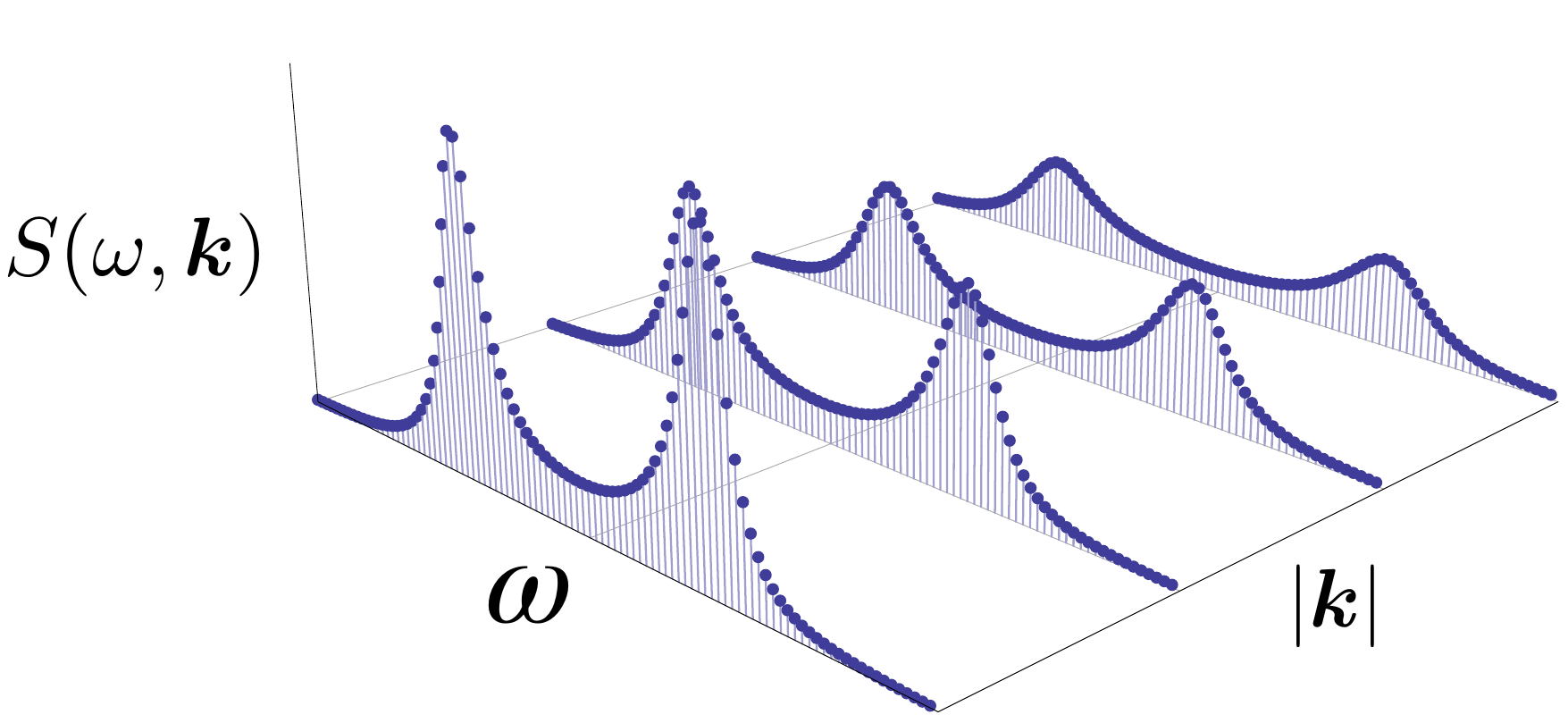}
        \end{center}
      \end{minipage} 
      \begin{minipage}{0.5\hsize}
        \begin{center}
          \includegraphics[width=\hsize]{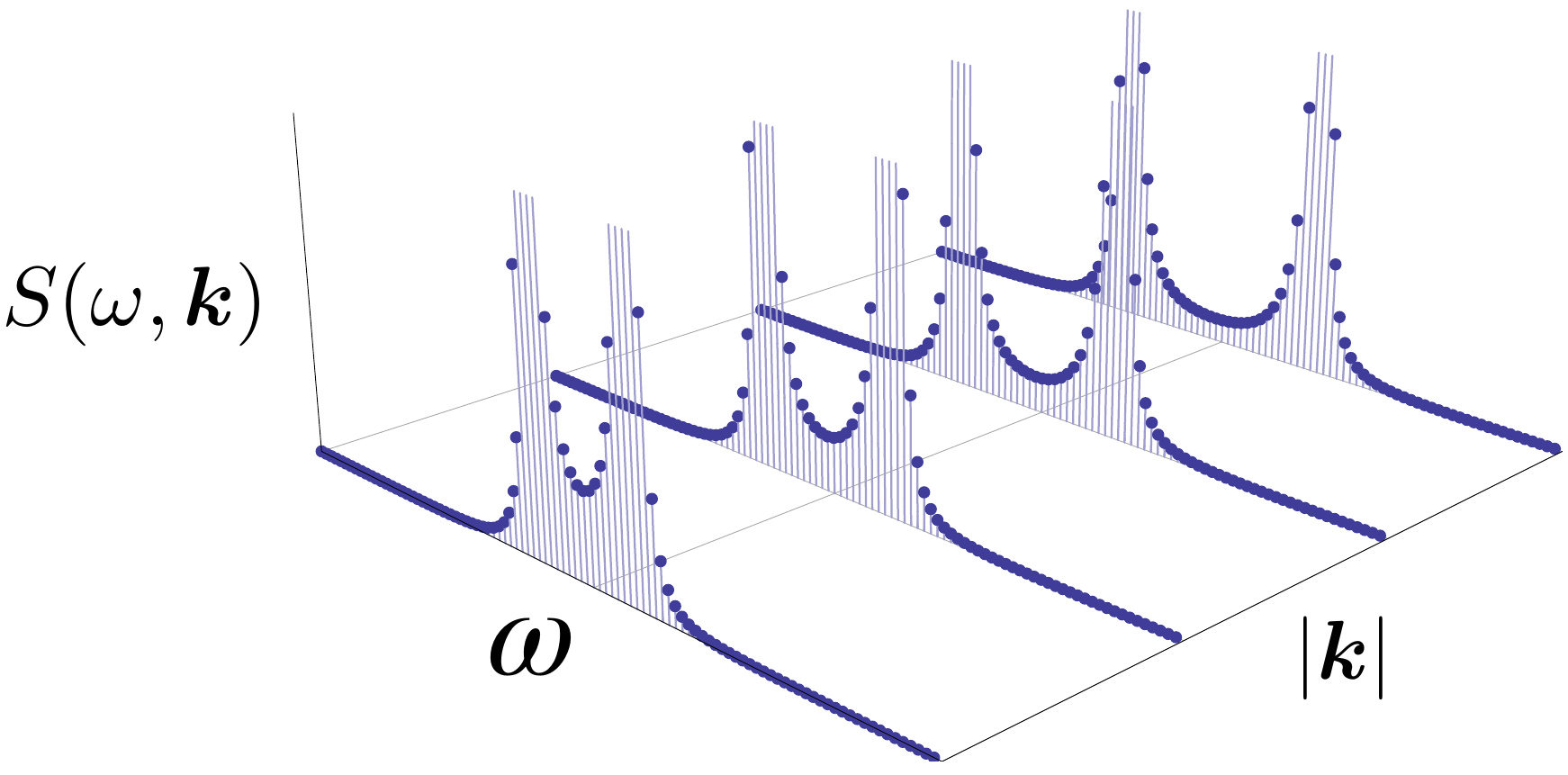}
        \end{center}
      \end{minipage}
    \end{tabular}
    \caption{The left and right panels show the $\omega$-$|\bm{k}|$ dependence of $S(\omega,\bm{k})$ for type-A and B modes in an isolated system, respectively. The both have the sharp pair peaks.  The parameters are chosen as $a_{A}=1$,  $b_{A}=0.5$, $a_{B}=1$, and $b_{B}=0.5$.}
    \label{fig:SpectralIsolated}
  \end{center}
\end{figure}
Figures~\ref{fig:SpectralOpen} and \ref{fig:SpectralIsolated} illustrate the $\omega$-$|\bm{k}|$ dependence of $S(\omega,\bm{k})$ in  open and isolated systems, respectively.
$S(\omega,\bm{k})$ in the open system has a single peak for the type-B diffusion mode, and blunt pair peaks for the type-B propagation mode.
In contrast,  $S(\omega,\bm{k})$ in the isolated system has sharp pair peaks.
Thus, $S(\omega,\bm{k})$ has significantly different behaviors depending on the open or the isolated system.

The type-B diffusion spectrum will be realized in a driven dissipative Bose--Einstein condensate (BEC)~\cite{PhysRevB.74.245316,PhysRevLett.99.140402,PhysRevA.76.043807,PhysRevLett.96.230602,Sieberer:2015svu}.
Similarly, we expect that the type-B propagation mode in a nonequilibrium steady state can be observed in a driven dissipative BEC with a different symmetry breaking pattern, e.g. $SO(3)\times U(1)\to U(1)$ realized in a spinor BEC~\cite{RevModPhys.85.1191, KAWAGUCHI2012253}.

\section{Open classical and quantum systems}\label{sec:OCQS}
Here, we briefly review the path integral approach to open classical and quantum systems. 
Readers who are familiar with this approach may jump to Sec.~\ref{sec:technique}.
As will be seen below, in both classical and quantum systems, the expectation value of an operator can be expressed as the path integral
\begin{equation}
\begin{split}
\langle \mathcal{O}[\chi^{a}_{R},\chi_{A}^{a}] \rangle =\int_{\rho} \mathcal{D} \chi_A^a   \mathcal{D} \chi_R^a
\mathcal{D} C   \mathcal{D} \bar{C}
 e^{\ri  S[\chi_R^a,\chi_{A}^{a},C,\bar{C}]}\mathcal{O}[\chi_{R}^{a},\chi_{A}^{a}].
 \label{eq:PathIntegral}
\end{split}
\end{equation}
Here, the subscript $\rho$ denotes the contribution from the initial density operator, whose definition is given later.
$S[\chi_R^a,\chi_{A}^{a},C,\bar{C}]$ is the action with two types of physical degrees of freedom, $\chi_R^a$ and $\chi_{A}^{a}$,
corresponding to classical and fluctuation fields, respectively. In classical systems, $\chi_{A}^{a}$ is often called the response field.
In quantum systems,  $\chi_R^a$ and $\chi_{A}^{a}$ are fundamental fields of the Keldysh basis in the Keldysh path formalism~\cite{keldysh1965diagram}. 
$C$ and $\bar{C}$ are ghost fields, which are responsible for the conservation of probability.
The existence of the ghost term depends on the theory.
In the following, we show three examples for classical and quantum systems that can be expressed as Eq.~\eqref{eq:PathIntegral}.

\subsection{Stochastic system}
The first example is a stochastic system. The path integral formalism in this system is the MSRJD formalism~\cite{MSR, J, D1, D2}. 
Let us consider a stochastic system whose dynamics is described by the Langevin type equation:
\begin{equation}
\begin{split}
\partial_{t}\phi =  -\gamma \phi -\lambda\phi^3+ \xi,
\label{eq:Langevin}
\end{split}
\end{equation}
where $\xi$ represents Gaussian white noise that satisfies 
\begin{equation}
\begin{split}
\langle \xi(t)\xi(t')\rangle_{\xi}=\kappa\delta(t-t').
\end{split}
\end{equation}
Here, $\kappa$ represents the strength of the noise.
The probability distribution is defined as 
\begin{equation}
\begin{split}
P[t;\phi_{R}] := \int d\phi_{R}(t_{I})\rho[\phi_{R}(t_{I})] \langle \delta(\phi_{R}- \phi(t))\rangle_{\xi},
\end{split}
\end{equation}
where $\rho[\phi_{R}(t_{I})]$ is the initial probability distribution.
Since $\phi(t)$ follows Eq.~\eqref{eq:Langevin},  $P[t_{F};\phi_{R}]$ can be expressed as
\begin{equation}
\begin{split}
P[t_{F};\phi_{R}] = \int_{\rho,\phi_{R}}\mathcal{D}\phi_{R} \det(\partial_{t}+ \gamma+3\lambda\phi_{R}^2) \langle \delta(\partial_{t}\phi_{R} +\gamma \phi_{R}+\lambda\phi_{R}^3 -{\xi})\rangle_{\xi},
\end{split}
\end{equation}
where $\det(\partial_{t}+\gamma+3\lambda\phi_{R}^2) $ is the Jacobian associated with the transformation from $(\phi_{R}-\phi(t))$ to $(\partial_{t}\phi_{R}+\gamma\phi_{R}+\lambda\phi_{R}^{3}-\xi(t))$. 
Here, we introduced the following notation:
\begin{equation}
\begin{split}
\int_{\rho,\phi_{R}}\mathcal{D}\phi_{R}:= \int_{\phi_{R}(t_{F})=\phi_{R}} \mathcal{D}\phi_{R} \rho[\phi_{R}(t_{I})].
\end{split}
\end{equation}
Using the Fourier representation of the delta function, we can write the probability distribution as
\begin{equation}
\begin{split}
P[t_{F};\phi_{R}] = \int_{\rho,\phi_{R}}\mathcal{D}\phi_{A} \mathcal{D}\phi_{R}  \det(\partial_{t}+\gamma+3\lambda\phi_{R}^2) \Bigl\langle \exp  \int_{t_{I}}^{t_{F}} dt\, \ri \phi_{A}(\partial_{t}\phi_{R} +\gamma \phi_{R}+\lambda\phi_{R}^3-{\xi})\Bigr\rangle_{\xi}.
\end{split}
\end{equation}
The average of the noise can be evaluated as
\begin{equation}
\begin{split}
\langle e^{-\ri\int dt \phi_{A}(t)\xi(t)}\rangle_{\xi} = e^{-\frac{\kappa}{2}\int dt\phi^{2}_{A}(t)},
\end{split}
\end{equation}
which leads to 
\begin{equation}
\begin{split}
P[t_{F};\phi_{R}] = \int_{\rho,\phi_{R}} \mathcal{D}\phi_{A}\mathcal{D}\phi_{R}\det(\partial_{t}+\gamma+3\lambda\phi_{R}^2) \exp \int_{t_{I}}^{t_{F}} dt\Bigl[ \ri\phi_{A}(\partial_{t}\phi_{R} +\gamma\phi_{R}+\lambda\phi_{R}^3)- \frac{\kappa}{2}\phi_{A}^{2}\Bigr].
\end{split}
\end{equation}
The determinant can be expressed as the Fermionic path integral:
\begin{equation}
\begin{split}
\det(\partial_{t}+\gamma+3\lambda\phi_{R}^2) = \int \mathcal{D}C\mathcal{D}\bar{C} \exp{\int dt \bar{C}(\partial_{t}+\gamma+3\lambda\phi_{R}^2)C}.
\end{split}
\end{equation}
We eventually obtain the path-integral formula:
\begin{equation}
\begin{split}
P[t_{F};\phi_{R}] = \int_{\rho,\phi_{R}}\mathcal{D}\phi_{A}\mathcal{D}\phi_{R} \mathcal{D}C\mathcal{D}\bar{C} e^{\ri S[\phi_{R},\phi_{A}, C,\bar{C}]},
\end{split}
\end{equation}
with the action, 
\begin{equation}
\begin{split}
 S[\phi_{R},{\phi}_{A}, C,\bar{C}] =  \int_{t_{I}}^{t_{F}} dt\Bigl[ {\phi}_{A}(\partial_{t}\phi_{R} +\gamma \phi_{R}+\lambda\phi_{R}^3)+\ri \frac{\kappa}{2}{\phi}_{A}^{2}\Bigr]
-\ri\int_{t_{I}}^{t_{F}} dt \bar{C}(\partial_{t}+\gamma+3\lambda\phi_{R}^2)C.\label{eq:MSRaction1}
\end{split}
\end{equation}
The expectation value of an operator $\mathcal{O}[\phi_{R},\phi_{A}]$ is given as
\begin{equation}
\begin{split}
\langle \mathcal{O}[\phi_{R},\phi_{A}] \rangle&:= 
\int d\phi_{R}(t_{F}) \int_{\rho,\phi_{R}}\mathcal{D}\phi_{A} \mathcal{D}\phi_{R}\mathcal{D}C\mathcal{D}\bar{C} e^{\ri S[\phi_{R},\phi_{A}, C,\bar{C}]}  \mathcal{O}[\phi_{R},\phi_{A}]\\
&=:\int_{\rho}\mathcal{D}\phi_{A}\mathcal{D}\phi_{R} \mathcal{D}C\mathcal{D}\bar{C} e^{\ri S[\phi_{R},\phi_{A}, C,\bar{C}]}  \mathcal{O}[\phi_{R},\phi_{A}].
\end{split}
\end{equation}
In the second line, we include the integral of $\phi_{R}(t_{F})$ at the boundary into $\int_{\rho}\mathcal{D}\phi_{R}$.
Here, we only considered a single variable with  simple interaction and noise.
Generalization to multi-component fields is straightforward.
For more detailed derivation, see Refs.~\cite{justin1989quantum, altland2010condensed, TAUBER20127}.

\subsection{Open quantum systems}
\begin{figure}[htbp] 
   \centering
   \includegraphics[width=0.5\linewidth]{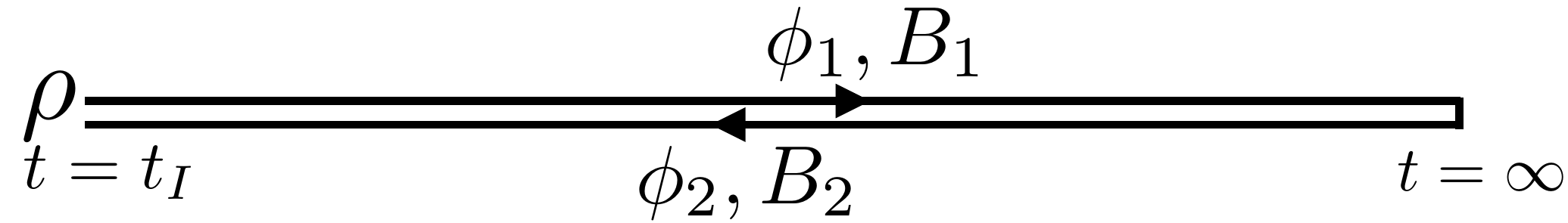} 
   \caption{Closed time contour in the Keldysh formalism.}
   \label{fig:path}
\end{figure}
The second example is an open quantum system. The open system can be formally constructed from an isolated system.
To see this, let us consider  a quantum system coupled with an environment. 
The path integral formula can be  obtained by integrating the environment out~\cite{Feynman:1963fq}.
The action of the total system consists of three parts:
\begin{equation}
\begin{split}
S_\text{tot}[\phi,B] = S_\text{sys}[\phi] + S_\text{env}[B] + S_\text{int}[\phi,B],
\end{split}
\end{equation}
where $S_\text{sys}[\phi]$, $S_\text{env}[B]$, and $S_\text{int}[B,\phi]$ are the actions of  the system, the environment, and the interaction between them, respectively.
$\phi$ and $B$ are the degrees of freedom of the system and the environment.
We assume that the initial density operator at $t=t_{I}$ is the direct product of those of the system and the environment: $\hat\rho= \hat\rho_\text{sys}\otimes \hat\rho_\text{env}$.
In the path integral formalism, the expectation value of an operator $\mathcal{O}[\phi_{1},\phi_{2}]$ is given on the path shown in Fig.~\ref{fig:path} as
\begin{equation}
\begin{split}
\langle \mathcal{O}[\phi_{1},\phi_{2}]\rangle &=  \int \mathcal{D}\phi_{1}\mathcal{D}\phi_{2}  \mathcal{D}B_{1}\mathcal{D}B_{2} \rho_\text{sys}[\phi_{1}(t_{I}),\phi_{2}(t_{I})]\rho_\text{env}[B_{1}(t_{I}),B_{2}(t_{I})]\\
&\quad\times e^{\ri S_\text{tot}[\phi_{1},B_{1}]-\ri S_\text{tot}[\phi_{2},B_{2}]}\mathcal{O}[\phi_{R},\phi_{A}],
\label{eq:SK}
\end{split}
\end{equation}
where the indices $1$ and $2$ represent the label of the forward and backward paths in Fig.~\ref{fig:path},
and we introduced the matrix elements of the density operators $\rho_\text{sys}[\phi_{1}(t_{I}),\phi_{2}(t_{I})]:=\langle \phi_{1}(t_{I})|\hat\rho_\text{sys}|\phi_{2}(t_{I}) \rangle$
and $\rho_\text{env}[B_{1}(t_{I}),B_{2}(t_{I})]:=\langle B_{1}(t_{I})|\hat\rho_\text{env}|B_{2}(t_{I}) \rangle$.
The direct product of the density operators enables us to formally  integrate the environment out.
Introducing the influence functional $e^{\ri\Gamma[\phi_{1},\phi_{2}]}$~\cite{Feynman:1963fq} defined as
\begin{equation}
\begin{split}
e^{\ri\Gamma[\phi_{1},\phi_{2}]}&:= \int   \mathcal{D}B_{1}\mathcal{D}B_{2}\rho_\text{env}[B_{I1},B_{I2}]
 e^{\ri S_\text{env}[B_{1}]-\ri S_\text{env}[B_{2}]+\ri S_\text{int}[\phi_{1},B_{1}]-\ri S_\text{int}[\phi_{2},B_{1}]},
\end{split}
\end{equation}
 we can express the expectation value as
\begin{equation}
\begin{split}
\langle \mathcal{O}[\phi_{1},\phi_{2}]\rangle=\int_{\rho} \mathcal{D}\phi_{1}\mathcal{D}\phi_{2} e^{\ri S_\text{eff}[\phi_{1},\phi_{2}]}\mathcal{O}[\phi_{1},\phi_{2}],
\end{split}
\end{equation}
where we defined
\begin{equation}
\begin{split}
S_\text{eff}[\phi_{1},\phi_{2}]:=S_\text{sys}[\phi_{1}]-S_\text{sys}[\phi_{2}]+\Gamma[\phi_{1},\phi_{2}],
\end{split}
\end{equation}
and the notation of the subscript $\rho$ by 
\begin{equation}
\begin{split}
\int_{\rho} \mathcal{D}\phi_{1}\mathcal{D}\phi_{2} := \int \mathcal{D}\phi_{1}\mathcal{D}\phi_{2}\rho_\text{sys}[\phi_{1}(t_{I}),\phi_{I2}(t_{I})].
\end{split}
\end{equation}
To see the connection between $S_\text{eff}$ and the action in Eq.~(\ref{eq:MSRaction1}) in the MSRJD formalism, let us move on to 
the Keldysh basis, which is defined as $\phi_{R}:=(\phi_{1}+\phi_{2})/2$ and $\phi_{A}:=(\phi_{1}-\phi_{2})$. $\phi_{R}$ and $\phi_{A}$ are called classical and quantum fields, respectively. 
Expanding the action with respect to $\phi_{A}$, and keeping $\phi_{A}$ up to quadratic order, we can obtain the MSRJD action.
For example, we consider $\phi^4$ theory in $(0+1)$ dimension, whose action is given as 
\begin{equation}
\begin{split}
S_\text{sys}[\phi] = \int dt\Bigl[\frac{1}{2}(\partial_{t}\phi)^{2} -\frac{1}{2}m^2\phi^2- \frac{u}{4} \phi^4\Bigr], \label{eq:scalaraction}
\end{split}
\end{equation}
In the Keldysh basis, $S_\text{sys}[\phi_{1}]-S_\text{sys}[\phi_{2}]$ is expressed as
\begin{equation}
\begin{split}
&S_\text{sys}[\phi_{1}]-S_\text{sys}[\phi_{2}] 
=  \int dt \Bigl[
\partial_{t}\phi_{A}\partial_{t}\phi_{R}-m^2\phi_{A}\phi_{R}-u\phi_{A}\phi_{R}^3-\frac{u}{4}\phi_{R}\phi_{A}^3
\Bigr].
\label{eq:systemaction}
\end{split}
\end{equation}
The functional form of $\Gamma[\phi_{1},\phi_{2}]$ depends on the details of the environment. 
For a simple case, we assume
\begin{equation}
\begin{split}
\Gamma[\phi_{1},\phi_{2}] = \int dt \Bigl[
-\nu \phi_{A} \partial_{t}\phi_{R} +  \frac{\ri}{2}D \phi_{A}^{2}
\Bigr], 
\label{eq:Gamma}
\end{split}
\end{equation}
where $\nu$ and $D$ are parameters coming from couplings to the environment.
Then, the action reduces to
\begin{align}
S_\text{eff}&=\int dt \Bigl[
\partial_{t}\phi_{A}\partial_{t}\phi_{R}-\nu\phi_{A}\partial_t \phi_{R}-m^2\phi_{A}\phi_{R}
-u\phi_{A}\phi_{R}^3-\frac{u}{4}\phi_{R}\phi_{A}^3
+ \frac{\ri}{2} D \phi_{A}^{2}
\Bigr]. 
\label{eq:SKaction0}
\end{align}
For a classical treatment, we drop $\phi_{A}^3$ term, and neglect the $\partial_{t}\phi_{A}\partial_{t}\phi_{R}$ term for slow dynamics.
Then, we arrive at
\begin{align}
 S_\text{eff}&=\int dt \Bigl[\phi_{A}(\partial_t \phi_{R}+\gamma\phi_{R}
+\lambda\phi_{R}^3)
+ \frac{\ri}{2} \kappa \phi_{A}^{2}
\Bigr], \label{eq:SKaction1}
\end{align}
where we rescaled $\phi_{A}\to - \phi_{A}/\nu$ and introduced $\gamma = m^2/\nu$, $\kappa=D/\nu^{2}$ and $\lambda=u/\nu$.
We can see that the action in Eq.~(\ref{eq:SKaction1}) is the same form as the MSR one in Eq.~(\ref{eq:MSRaction1}) except for the Jacobian term.

\subsection{Lindblad equation}
The third example is a system described by a Lindblad equation~\cite{Lindblad:1975ef} (see Ref.~\cite{Sieberer:2015svu} for a review of the path integral formalism of the Lindblad equation).  The Lindblad equation with a single Lindblad operator $\hat{L}$ is given as
\begin{equation}
\begin{split}
\partial_{t}\hat{\rho} = -\ri [\hat{H},\hat{\rho}]+ \gamma \Bigl(\hat{L}\hat{\rho} \hat{L}^{\dag}-\frac{1}{2}(\hat{L}^{\dag}\hat{L}\hat{\rho} +\hat{\rho} \hat{L}^{\dag}\hat{L}) \Bigr).
\end{split}
\end{equation}
Here, $\gamma$ is a coefficient and $\hat{L}$ is a function of fields.  The term proportional to $\gamma$ describes the dissipation and fluctuation effects.
By construction, the trace of $\hat{\rho}$ is time independent, $\partial_{t}\tr \hat{\rho}=0$.

When the degrees of freedom is a bosonic Schr\"odinger field $\hat{\psi}(t,\bm{x})$ in $(d+1)$ dimensions with the commutation relation $[\hat\psi(t,\bm{x}),\hat\psi^{\dag}(t,\bm{x}')] = \delta^{(d)}(\bm{x}-\bm{x}')$, the path integral for the expectation value of an operator $\mathcal{O}[\psi_{i},\psi^{\dag}_{i}]$ is given as
\begin{equation}
\begin{split}
\langle \mathcal{O}[\psi_{i},\psi^{\dag}_{i}]\rangle =\int_{\rho} \mathcal{D}\psi_{i}\mathcal{D}\psi_{i}^{\dag}e^{\ri S} \mathcal{O}[\psi_{i},\psi^{\dag}_{i}]
\end{split}
\end{equation}
with the action,
\begin{equation}
\begin{split}
S &=  \int d^{d+1}x\, \Bigl[ \psi_{1}^{\dag}\ri \partial_{t} \psi_{1}- H_{1}-\psi_{2}^{\dag}\ri \partial_{t} \psi_{2}  +H_{2}
-\ri \gamma\Bigl( L_{1}L_{2}^{\dag} -\frac{1}{2}( L_{1}^{\dag}L_{1} + L^{\dag}_{2}L_{2} ) \Bigr)
\Bigr],
\end{split}
\end{equation}
where $H_{i}:= \langle \psi_{i}|\hat{H}|\psi_{i}\rangle$, and $L_{i}=\langle\psi_{i}|\hat{L}|\psi_{i}\rangle$,  $L^{\dag}_{i}=\langle\psi_{i}|\hat{L}^{\dag}|\psi_{i}\rangle$~\footnote{
Precisely, speaking, when $\hat{L}$ depends on both $\hat{\psi}$ and $\hat{\psi}^{\dag}$, one need to take care of the ordering of the field in $\hat{L}^{\dag}\hat{L}$, since $\hat{\psi}$ and $\hat{\psi}^{\dag}$ are not commutative.}.
For example, if we choose the Hamiltonian and the Lindblad operator as
\begin{equation}
\begin{split}
\hat{H} = \frac{1}{2m}\bm{\nabla}\hat{\psi}^{\dag}\bm{\nabla}\hat{\psi} ,\qquad \hat{L}=\sqrt{2}\hat{\psi},
\end{split}
\end{equation}
the action is written in the form,
\begin{equation}
\begin{split}
S &=  \int d^{d+1}x\, \Bigl[ \psi_{1}^{\dag}\ri \partial_{t} \psi_{1}- \frac{1}{2m}\bm{\nabla}\psi_{1}^{\dag}\bm{\nabla}\psi_{1}-\psi_{2}^{\dag}\ri \partial_{t} \psi_{2}  + \frac{1}{2m}\bm{\nabla}\psi_{2}^{\dag}\bm{\nabla}\psi_{2}\\
&\qquad\qquad\quad-\ri\gamma\Bigl( 2\psi_{1}\psi_{2}^{\dag} -( \psi_{1}^{\dag}\psi_{1} + \psi_{2}^{\dag}\psi_{2} ) \Bigr)
\Bigr],
\label{eq:LindbladAction}
\end{split}
\end{equation}
where $m$ is the mass. In the Keldysh basis, it becomes
\begin{equation}
\begin{split}
S &=  \int d^{d+1}x\, \Bigl[ \psi_{R}^{\dag}\ri (\partial_{t}-\gamma )\psi_{A}+\psi_{A}^{\dag}\ri (\partial_{t} +\gamma)\psi_{R}- \frac{1}{2m}\bm{\nabla}\psi_{R}^{\dag}\bm{\nabla}\psi_{A}
- \frac{1}{2m}\bm{\nabla}\psi_{A}^{\dag}\bm{\nabla}\psi_{R}
+\ri\gamma\psi^{\dag}_{A}\psi_{A}
\Bigr].
\end{split}
\end{equation}

As we have seen in the three examples, all the systems are described by the path integral. 

\section{Field-theoretical technique}\label{sec:technique}
In this section, we provide the concept of symmetry in open systems and two field-theoretical techniques to derive low-energy coefficients. 
One is the Ward--Takahashi identity that gives the relation between different Green functions.
The other is the generating functional and effective action methods, which are often employed to prove the Nambu--Goldstone theorem in quantum field theories (see, e.g., Ref.~\cite{WeinbergText}).
\subsection{Symmetry of open systems}\label{sec:symmetry}
In isolated systems, the existence of a continuous symmetry implies the existence of  conserved current. 
On the other hand, in open systems the energy, momentum, and particle number are generally not conserved due to the interaction between the system and the environment. Thus, one may wonder what  symmetry of open systems means. Even in such a case, symmetry may exist ~\cite{Sieberer:2015svu,Minami:2018oxl}.  
To see the symmetry of open systems, let us consider the concrete example with the action in Eq.~\eqref{eq:LindbladAction}. 
If $\gamma=0$, this system corresponds to an isolated system. In this case, the action is invariant under $\psi_{1}\to e^{\ri\theta_{1}}\psi_{1}$ and  $\psi_{2}\to e^{\ri\theta_{2}}\psi_{2}$, where $\theta_{1}$ and $\theta_{2}$ are constant.  In this sense, there is the $U(1)_{1}\times U(1)_{2}$ symmetry, whose Noether charges are 
\begin{equation}
\begin{split}
Q_{1} = \int d^{d}x \psi_{1}^{\dag}\psi_{1}, \qquad Q_{2} = \int d^{d}x \psi_{2}^{\dag}\psi_{2},
\end{split}
\end{equation}
respectively. Here, we defined $Q_{2}$ such that $[\hat{Q}_{2}, \hat{\psi}_{2}]=+\hat{\psi}_{2}$ in the operator formalism, whose sign is opposite to $[\hat{Q}_{1},\hat{\psi}_{1}]= -\hat{\psi}_{1}$. This relative sign is caused by the canonical commutation relation, $[\hat{\psi}_{1}(t,\bm{x}), \hat{\psi}_{1}^{\dag}(t,\bm{x}')]=\delta^{(d)}(\bm{x}-\bm{x}')$ and $[\hat{\psi}_{2}(t,\bm{x}), \hat{\psi}_{2}^{\dag}(t,\bm{x}')]=-\delta^{(d)}(\bm{x}-\bm{x}')$ in the operator formalism. This can be understood as the sign in front of the time derivative term in the Keldysh action $\int d^{d+1} (\psi_{1}^{\dag}\ri \partial_{t} \psi_{1}-\psi_{2}^{\dag}\ri \partial_{t} \psi_{2})$. 
This kind of doubled symmetry is used for the construction of an effective field theory of fluids in the Keldysh formalism~\cite{Harder:2015nxa,Haehl:2015pja,Crossley:2015evo,Glorioso:2016gsa,Glorioso:2017fpd,Jensen:2017kzi,Jensen:2018hhx,Haehl:2016pec,Geracie:2017uku,Haehl:2018uqv,Haehl:2018lcu,Glorioso:2018wxw}.
In the Keldysh basis, these are expressed as
\begin{align}
Q_{A} &:= Q_{1}-Q_{2}=\int d^{d}x(\psi^{\dag}_{R}\psi_{A}+\psi^{\dag}_{A}\psi_{R}),\\
Q_{R} &:= \frac{1}{2}(Q_{1}+Q_{2})=\int d^{d}x\Bigl(\psi^{\dag}_{R}\psi_{R}+\frac{1}{4}\psi^{\dag}_{A}\psi_{A}\Bigr).
\end{align}
In isolated systems, these charges generate an infinitesimal transformation as
\begin{align}
\delta_{A}\psi_{A}&= \ri \psi_{A},\qquad  \delta_{A}\psi_{R} = \ri \psi_{R}, \label{eq:transformationA}\\
\delta_{R}\psi_{A} &= \ri \psi_{R},\qquad  \delta_{R}\psi_{R} = \frac{1}{4}\ri \psi_{A}, \label{eq:transformationR}
\end{align} 
where $\delta_{i}\hat{\psi}_{j}=-\ri [\hat{Q}_{i}, \hat{\psi}_{j}]$ in the operator formalism.
On the other hand, when $\gamma\neq0$, one of the symmetries is explicitly broken. The residual symmetry is $U(1)_{A}$, which is given by setting the parameters as $\theta_{1}=\theta_{2}$. The Noether charge $Q_{A}$ is still conserved, but $Q_{R}$ is not, which means that the charge in the usual sense is not conserved.

More generally, we can consider a system with continuous symmetry $G$. 
The symmetry of open systems means that the action is invariant under  the infinitesimal transformation of $G$, 
\begin{equation}
\begin{split}
\chi_{A}^{a}(x)\to \chi_{A}^{a}(x)+\epsilon_{\alpha}\delta_{A}^{\alpha}\chi_{A}^{a}(x),\qquad
\chi_{R}^{a}(x)\to \chi_{R}^{a}(x)+\epsilon_{\alpha}\delta_{A}^{\alpha}\chi_{R}^{a}(x),
\end{split}
\end{equation}
where $\chi^{a}_{A}(x)$ and $\chi^{a}_{R}(x)$ are field degrees of freedom, and $\epsilon_{\alpha}$ is constant.
We can formally define the transformation $\delta_{R}^{\alpha}\chi_{i}(x)$ such that 
\begin{equation*}
\begin{split}
\delta_{R}^{\alpha}\chi_{A}^{a}(x) &:= \delta_{A}^{\alpha}\chi_{R}^{a}(x),\\
\delta_{R}^{\alpha}\chi_{R}^{a}(x)&: = 
\begin{cases}
 \frac{1}{4}\delta_{A}^{\alpha}\chi_{A}^{a} (x)& \text{quantum system}\\
 0 &\text{classical system}
 \end{cases}.
 \end{split}
 \tag{\ref{eq:transR}}
\end{equation*}
These are generalizations of Eqs.~\eqref{eq:transformationA} and \eqref{eq:transformationR}~\cite{Minami:2018oxl}.  
In open systems, invariance of the action under $\delta_{R}^{\alpha}$ depends on the theory.
If the particle number is conserved, but energy is not, then, the action is invariant under the $U(1)_{R}$ transformation
$\chi_{A}\to  e^{\ri\theta}\chi_{R}$ and $\chi_{R}\to  e^{\ri\theta}\chi_{A}/4$, 
but it is not invariant under the time translation $\chi_{A}(t)\to  \chi_{R}(t+c)$ and $\chi_{R}(t)\to  \chi_{A}(t+c)/4$,
where $\theta$ and $c$ are constants.

In addition, all the previous examples in Sec.~\ref{sec:OCQS} satisfy the reality condition:
\begin{equation}
\begin{split}
\bigl(S[\chi^{a}_{R},\chi^{a}_{A} ] \bigr)^{*} =  - S[\chi^{a}_{R},-\chi^{a}_{A} ].
\label{eq:RealityCondition}
\end{split}
\end{equation}
In this paper, we focus on theories satisfying the reality condition.

\subsection{Ward--Takahashi identity}\label{sec:WardTakahashi}
Symmetry plays an important role not only in isolated systems but also in open ones. 
Here, we show the Noether current in open systems and the Ward--Takahashi identity~\cite{PhysRev.78.182,Takahashi1957}. 
Let us consider a theory with field variables $\chi_A^{a}$ and $\chi_{R}^{a}$.
Suppose the action $S$ is invariant under $G$ and the infinitesimal transformation is given as $\chi^{a}_{i}\to \chi^{a}_{i}+\epsilon_{\alpha}\delta^{\alpha}_{A} \chi^{a}_{i}$,
with an infinitesimal constant $\epsilon_{\alpha}$. When $\epsilon_{\alpha}(x)$ depends on the spacetime coordinate,  the action transforms as
\begin{equation*}
\delta_{A} S =  -\int d^{d+1}x\, j_{A}^{\alpha\mu}(x) \partial_{\mu}\epsilon_{\alpha}(x),
\tag{\ref{eq:deltaAS}}
\end{equation*}
because $\delta_{A} S$ vanishes if $\epsilon_{\alpha}(x)$ is constant.
The operator $ j_{A}^{\alpha\mu}(x)$ is  called the Noether current. To see the Ward--Takahashi identity in the path integral formalism, we consider the expectation value of an operator $\mathcal{O}[\chi^{a}_{i}]$, which is given as
\begin{equation}
\begin{split}
\langle \mathcal{O}[\chi^{a}_{i}]\rangle =\int \mathcal{D}\chi^{a}_{i} e^{\ri S[\chi^{a}]} \mathcal{O}[\chi^{a}_{i}]  .
\end{split}
\end{equation}
Since $\chi^{a}_{i}$ is the integral variable, the integral is invariant under the relabeling $\chi^{a}_{i}\to \chi'^{a}_{i}$:
\begin{equation}
\begin{split}
\langle \mathcal{O}[\chi^{a}_{i}]\rangle =\int \mathcal{D}\chi'^{a}_{i} e^{\ri S[\chi'^{a}]} \mathcal{O}[\chi'^{a}_{i}]  .
\end{split}
\end{equation}
If we choose $\chi'^{a}_{i}(x)= \chi^{a}_{i}(x)+ \epsilon_{\alpha}(x)\delta_{A}^{\alpha}\chi^{a}_{i}(x)$, and assuming that  the path-integral measure is invariant under this transformation: $\mathcal{D}\chi'^{a}_{i}  = \mathcal{D}\chi^{a}_{i} $, we find that
\begin{equation}
\begin{split}
\langle \mathcal{O}[\chi^{a}_{i}]\rangle &=\int \mathcal{D}\chi^{a}_{i} e^{\ri S[\chi^{a}]+\ri\delta_{A} S } (\mathcal{O}[\chi^{a}_{i}]+ \delta_{A} \mathcal{O}[\chi^{a}_{i}]) +O(\epsilon^{2})\\
&= \langle \mathcal{O}[\chi^{a}_{i}]\rangle +  \langle \delta_{A} \mathcal{O}[\chi^{a}_{i}]\rangle+\ri\langle \delta_{A} S\, \mathcal{O}[\chi^{a}_{i}]\rangle 
+O(\epsilon^{2}).
\end{split}
\end{equation}
Here, the local transformation of $\mathcal{O}[\chi^{a}_{i}]$ is given as
\begin{equation}
\begin{split}
 \delta_{A} \mathcal{O}[\chi^{a}_{i}]=\int d^{d+1}x\,\epsilon_{\alpha}(x)\frac{\delta\mathcal{O}[\chi^{a}_{i}]}{\delta \chi^{b}_{j}(x)}\delta^{\alpha}_{A}\chi^{b}_{j}(x).
\end{split}
\end{equation}
The leading-order term in $\epsilon_{\alpha}(x)$ gives 
\begin{equation}
\begin{split}
\Bigl\langle \int d^{d+1}x\, j_{A}^{\alpha\mu}(x) \partial_{\mu}\epsilon_{\alpha}(x) \mathcal{O}[\chi^{a}_{i}]\Bigr \rangle +\ri\langle \delta_{A} \mathcal{O}[\chi^{a}_{i}]\rangle =0.
\label{eq:wardid0}
\end{split}
\end{equation}
Differentiating Eq.~\eqref{eq:wardid0} with respect to $\epsilon_{\alpha}(x)$, we obtain
\begin{equation}
\begin{split}
-\partial_{\mu} \langle j_{A}^{\mu}(x)  \mathcal{O}[\chi^{a}_{i}] \rangle  +\ri \Bigl\langle \frac{\delta \mathcal{O}[\chi^{a}_{i}]}{\delta \chi^{a}_{i}(x)}\delta_{A}^{\alpha}\delta\chi^{a}_{i}(x)\Bigr\rangle=0.
\end{split}
\end{equation}
This is the Ward--Takahashi identity, which is the conservation law in the path integral formalism.

\subsection{Generating functional and effective action}\label{sec:EffectiveAction}
In order to show the Nambu--Goldstone theorem, it is useful to introduce the generating functional.
We start with a path integral representation for the generating functional in $(d+1)$ spacetime dimensions:
\begin{align}
Z[J]:=\Bigl\langle \exp\Bigl[\ri \int {d^{d+1}x}\,  J^i_a(x)\phi_{i}^a(x) \Bigr] \Bigr\rangle  = \int_{\rho} \mathcal{D} \chi_i^a \exp\Bigl[\ri  S[\chi_i^a]+\ri \int {d^{d+1}x}\,  J^i_a(x)\phi_{i}^a(x) \Bigr], 
\label{eq:path-integral}
\end{align} 
where the $\chi_i^a$ are elementary degrees of freedom in the Keldysh basis. $\phi_{i}^a=(\phi_{R}^a,\phi_{A}^a)$ is a set of real scalar fields, which may be the elementary or a composite field of $\chi_i^a$\footnote{When $\phi_{R}^a$ and $\phi_{A}^a$ are composite, these fields are defined by using $1/2$ basis such that $\phi_{R}=(\phi_{1}+\phi_{2})/2$ and $\phi_{A}=\phi_{1}-\phi_{2}$, where $\phi_{1}$ and $\phi_{2}$ are polynomials of $\chi_{1}$ and $\chi_{2}$, respectively.}.
In general, the stationary state need not be the thermal equilibrium state, i.e., a nonequilibrium steady state is allowed in this formalism. For the generating functional to be well defined,  we assume that the stationary state is stable against any small perturbations. 
We also assume a stationary state that is independent of the choice of the initial density operator, so that we omit the subscript $\rho$ in the following.

Connected Green functions are generated by differentiating  $\ln Z[J]$ with respect to $J$: 
\begin{equation}
\begin{split}
\langle \phi_{i_{1}}^{a_{1}}(x_{1})\cdots \phi_{i_{n}}^{a_{n}}(x_{n}) \rangle_{\text{c};J} = \frac{1}{\ri^{n}}\frac{\delta^{n} \ln Z[J]}{\delta J_{a_{1}}^{i_{1} }(x_{1})\cdots\delta J_{a_{n}}^{i_{n} }(x_{n})}.
\end{split}
\end{equation}
Here, $\langle ... \rangle_J$ denotes the expectation value in the presence of the source ${J}_a^i(x)$, which is defined as
\begin{align}
\langle \mathcal{O} \rangle_J  := \frac{1}{Z[{J}]}\int \mathcal{D} {\chi}_i^ae^{\ri  S[{\chi}]+\ri \int {d^{d+1}x}\,  {J}^i_a(x){\phi}_i^a(x)} \mathcal{O}.
\end{align}
The subscript $c$ denotes the connected part of the correlation function.
The Green function without the external field is given as the limit of $J_{a}^{i}(x)\to 0$:
\begin{equation}
\begin{split}
\langle \phi_{i_{1}}^{a_{1}}(x_{1})\cdots \phi_{i_{n}}^{a_{n}}(x_{n}) \rangle_\text{c} = 
\lim_{J\to0}\langle \phi_{i_{1}}^{a_{1}}(x_{1})\cdots \phi_{i_{n}}^{a_{n}}(x_{n}) \rangle_{\text{c};J}.
\end{split}
\end{equation}

Since the $J_{a}^{A}\phi^{a}_{A}$ term in Eq.~\eqref{eq:path-integral} is proportional to $\phi^{a}_{A}$, it can be absorbed into the action as the external force, while $J_{a}^{R}\phi^{a}_{R}$ cannot. The external force does not change the conservation of the probability, so that the partition function becomes trivial  if $J^{R}=0$: $Z[J^R=0, J^A]=1$.
This identity leads to 
\begin{equation}
\begin{split}
\langle \phi_{A}^{a_{1}}(x_{1})\cdots \phi_{A}^{a_{n}}(x_{n}) \rangle_\text{c} = \left.\frac{1}{\ri^{n}}\frac{\delta^{n} \ln Z[J]}{\delta J_{a_{1}}^{A }(x_{1})\cdots\delta J_{a_{n}}^{A}(x_{n})}\right|_{J=0}=0.
\label{eq:AAAA}
\end{split}
\end{equation}
Therefore, any correlation functions contracted from only the $\phi_{A}$ vanish. In particular, $\langle \phi_{A}^{a}(x)\rangle$ cannot be the order parameter.

We introduce here the effective action, which is defined as 
\begin{align}
\Gamma [{\varphi}] := -\ri  \ln Z[{J}] - \int {d^{d+1}x}\, J^i_a(x) {\varphi}_i^a(x),
\end{align}
where $\varphi_i^a(x) := \langle {\phi}_i^a(x) \rangle_J =-\ri\delta \ln Z[J]/\delta J^i_a(x)$. 
We assume that $\varphi_i^a(x)=\langle {\phi}_i^a(x) \rangle_J$ is invertible so that the effective action is well defined.
The effective action is a functional of $\varphi_i^a(x)$ rather than $J^i_a(x)$,
since $\delta \Gamma [{\varphi}] =-\int d^{d+1}x\,\ri \delta J^i_a(x) \delta \ln Z[J]/\delta J^i_a(x) -\int d^{d+1}x\, \delta J_a^i(x)\varphi_i^a(x)-\int d^{d+1}x\, J_a^i(x)\delta\varphi_i^a(x)=-\int d^{d+1}x\, J_a^i(x)\delta\varphi_i^a(x)$, in which  $J_a^i(x)$ is not an independent variable but relates to $\varphi_i^a(x)$ through $\varphi_i^a(x) = \langle {\phi}_i^a(x) \rangle_J$.
The functional derivative of $\Gamma$ with respect to $\varphi_{i}^{a}$ is 
\begin{equation}
\begin{split}
\frac{\delta \Gamma[\varphi]}{\delta \varphi^{a}_{i}(x)} = -J_{a}^{i}(x).
\label{eq:GammaEOM}
\end{split}
\end{equation}
In the absence of the external field, $J_{a}^{i}(x)=0$, Eq.~\eqref{eq:GammaEOM} gives the stationary condition of the effective action.  
The second functional derivative of  $\Gamma$ is
\begin{equation}
\begin{split}
\frac{\delta^{2} \Gamma[\varphi]}{\delta \varphi^{b}_{j}(x')\delta \varphi^{a}_{i}(x)} = -\frac{\delta J_{a}^{i}(x)}{\delta \varphi_{j}^{b}(x')}
=-\left(\frac{\delta \varphi_{j}^{b}(x')}{\delta J_{a}^{i}(x)}\right)^{-1}
=-[D^{-1}]_{ba}^{ji}(x',x;\varphi).
\label{eq:inverserD}
\end{split}
\end{equation}
Here, we defined the two-point Green function as
\begin{equation}
\begin{split}
-\ri D_{ij}^{ab}(x,x';\varphi) := \langle \varphi^{a}_{i}(x)\varphi^{b}_{j}(x')\rangle_{c;J}  =-\ri \frac{\delta \varphi_{j}^{b}(x')}{\delta J_{a}^{i}(x)}.
\end{split}
\end{equation}

To see the symmetry of the effective action, we apply the Ward--Takahashi identity in Eq.~\eqref{eq:wardid0}.
By choosing $\mathcal{O} =  \exp{\ri \int d^{d+1}J_{a}^{i}(x)\phi^{a}_{i}(x)}$ in Eq.~\eqref{eq:wardid0}, we obtain 
\begin{align}
\int d^{d+1}x\, \partial_\mu \epsilon_\alpha(x)\langle j_A^{\alpha\mu}(x)\rangle_J=  \int {d^{d+1}x}\,  \epsilon_\alpha(x) {J}_a^i(x) 
\langle \delta_A^\alpha {\phi}_i^a(x) \rangle_J, \label{eq:conservationLaw}
\end{align}
or equivalently,
\begin{align}
\int d^{d+1}x \,\partial_\mu \epsilon_\alpha(x)\langle j_A^{\alpha\mu}(x)\rangle_J=  -\int {d^{d+1}x}\,  \epsilon_\alpha(x) \frac{\delta \Gamma}{\delta \varphi^{a}_{i}(x)}
\langle \delta_A^\alpha {\phi}_i^a(x) \rangle_J, \label{eq:WTIdentity1}
\end{align}
where we used Eq.~\eqref{eq:GammaEOM}.
These equations play an essential role in the analysis of Nambu--Goldstone modes.

\section{Spontaneous symmetry breaking and the Nambu--Goldstone theorem}
\label{sec:SSB}

In this section, we nonperturbatively establish the Nambu--Goldstone theorem in open systems.
We will derive the formulae for the inverse of the retarded Green function for NG modes shown in Eqs.~\eqref{eq:invG}-\eqref{eq:C10formula}.
We consider an open system with elementary fields $\chi^{a}_{i}(x)$.
The system has spacetime translational invariance, and a continuous internal symmetry $G$.
The fields transform under an infinitesimal transformation of $G$: $\chi_{i}\to \chi_{i}+\epsilon_{\alpha}\delta_{A}^{\alpha} \chi^{a}_{i}(x)$ with
$\delta_{A}^{\alpha} \chi^{a}_{i}(x) =\ri[ T^{\alpha}]^{a}_{~b}\chi_{i}^{b}(x)$, where the $T^{\alpha}$ are  generators of $G$.
We assume that the order parameter belongs to a set of real fields $\{\phi_i^a\}$ that transforms under $G$ as
\begin{align}
\phi_i^a(x) &\rightarrow \phi_i^a(x)+ \epsilon_\alpha \delta_A^\alpha \phi_i^a(x)
\end{align}
with $\delta^\alpha_A\phi_i^a(x) := \ri [T^\alpha]^{a}_{~b} \phi_i^b(x)$.
In other words, the $\phi_i^a$ belong to a linear representation of $G$.
$\phi_i^{a}$ may be elementary or composite. 
 The representation of $\phi_{i}^{a}$ may be different from the $\chi_{i}^{a}$.
 For technical reasons, we assume that $\phi_i^{a}$ transforms under the local transformation:
\begin{equation}
\begin{split}
\delta_{A} \phi_i^{a}(x) = \epsilon_{\alpha}(x)\ri[ T^{\alpha}]^{a}_{~b}\phi_i^{b}(x).
\label{eq:localphi}
\end{split}
\end{equation}
This assumption will make the derivation of our formulae simple.
We assume that the continuous symmetry $G$ is spontaneously broken to its subgroup $H$. 
The symmetry breaking is characterized by the nonvanishing order parameter, $\delta_A^\alpha {{\bar{\varphi}}_i}^a:=\langle \delta^{\alpha}_{A}\phi_i^a(x) \rangle$.
We also assume that the translational symmetry is not broken, i.e., the order parameter is independent of time and space, which enables us to work in momentum space. 

Our procedure consists of  three steps: 
First, we show the existence of gapless excitations. Second, we will find the relation between the inverse of the retarded Green functions for $\{\phi_{i}^{a}\}$ and their low-energy coefficients. In general,  $\{\phi_{i}^{a}\}$ contains not only NG fields but also fields with gapped modes that we are not interested in.  Therefore, in the third step, we derive the inverse of the retarded Green functions with only NG fields by projecting out the contribution from gapped modes.

\subsection{Existence of gapless excitations} \label{sec:WTIdentity1}
The existence of gapless excitations can be shown by using the standard technique developed in quantum field theory~\cite{WeinbergText}.
In the generating functional method, the order parameter is given as $\delta_A^\alpha {{\bar{\varphi}}_i}^a := \lim_{J\to0}\langle \delta_A^\alpha {\phi}_i^a(x) \rangle_J$.
Taking $ \epsilon_\alpha(x)$ to be constant in Eq.~\eqref{eq:WTIdentity1}, we obtain
\begin{align}
\int {d^{d+1}x}\, \frac{\delta \Gamma}{\delta \varphi_i^a(x)}\delta_A^\alpha \varphi^a_i(x)=0.
 \label{eq:equality}
\end{align}
Differentiating Eq.~\eqref{eq:equality} with respect to $\varphi^b_j(x')$, we get
\begin{align}
\int {d^{d+1}x}\,  [D^{-1}]_{ba}^{j i}(x',x;\varphi) \delta_A^\alpha \varphi^a_i(x)
= -J_a^j(x')\ri[ T^\alpha]_{~b}^a, 
 \label{eq:equality2}
\end{align}
where we used $\delta_A^\alpha \varphi^a_i(x) = \ri [T^\alpha]^a_{~b}\varphi^b_i(x)$ and Eq.~\eqref{eq:inverserD}.
Taking the limit $J_{a}^{j}\to0$, 
we obtain
\begin{align}
\int {d^{d+1}x}\,  [D^{-1}]_{ba}^{jR} (x'-x)\delta_A^\alpha {{\bar{\varphi}}_R}^a&=0, 
\end{align}
where $[D^{-1}]_{ba}^{ji} (x'-x):= [D^{-1}]_{ba}^{ji}(x',x;\varphi=\bar{\varphi})$ with $\bar{\varphi}^a_i:=\lim_{J\to0}\langle \phi_i^a(x)\rangle_J$.
In momentum space, we obtain  
\begin{align}
[D^{-1}]_{ba}^{AR} (k=0)\delta_A^\alpha {{\bar{\varphi}}_R}^a=0. \label{eq:wardid}
\end{align}  
This identity represents the eigenvalue equation with the zero eigenvalue, whose eigenvectors  are $\delta_A^\alpha {\bar{\varphi}^a_R}$, and it
implies the existence of gapless excitations.  One can check that the independent number of $\delta_A^\alpha {{\bar{\varphi}}_R}^a$ equals $\dim(G/H)=:N_\text{BS}$. In an isolated system with Lorentz symmetry, $N_\text{BS}$ is equal to the number of gapless excitations~\cite{WeinbergText}. 
However, this is not the case in open systems. To obtain information on finite frequency and momentum, we need the further data shown in the following subsections.

\subsection{Low-energy coefficients}
In order to obtain finite momentum information, let us go back to Eq.~\eqref{eq:WTIdentity1} and consider its functional derivative  with respect to $\varphi^b_j(x')$:
\begin{equation}
\frac{\delta }{\delta \varphi^b_j(x')} \langle\delta_AS\rangle_J
=
-\int {d^{d+1}x}\,  [D^{-1}]_{ba}^{j i}(x',x;\varphi) 
\delta_A\varphi^a_i(x)
 -\epsilon_\alpha(x')J_{a}^{j}(x') \ri[ T^\alpha]_{~b}^a.
\label{eq:basic}
\end{equation}
Here, we employed Eq.~\eqref{eq:deltaAS} on the left-hand side 
and defined the local transformation,
\begin{equation}
\begin{split}
\delta_A\varphi^a_i(x):=\epsilon_{\alpha}(x)\delta_A^\alpha\varphi^a_i(x)=\epsilon_{\alpha}(x)\ri [T^{\alpha}]^{a}_{~b}\varphi_{i}^{b}(x),
\end{split}
\end{equation}
 to simplify the notation. This local transformation is obeyed from Eq.~\eqref{eq:localphi}.
 We can also introduce the local transformation $\delta_R\varphi_i^a(x)$ defined as
 \begin{equation}
 \begin{split}
 \delta_R\varphi_i^a(x):= \bar\epsilon_{\beta}(x) \delta_R^\beta\varphi_i^a(x)=\bar\epsilon_{\beta}(x) \ri[T^\beta]^a_{~b}\varepsilon_{i}^{~j}\varphi_j^b(x).
 \end{split}
 \end{equation}
 Here, we introduced the symbol $\varepsilon_{i}^{~j}$ for the transformation defined from $\delta^\alpha_R\chi_i^a(x) = \varepsilon_i^{~j} \delta_{A}^{\alpha}\chi_j^a(x)$ in Eq.~\eqref{eq:transR}.
The bar on the $\bar\epsilon_{\alpha}$ is introduced to distinguish the transformation for $\delta_{A}$, where $\epsilon_{\alpha}$ is used.
Multiplying Eq.~\eqref{eq:basic} by $-\int d^{d+1}x' \, \delta_R\varphi_j^b(x')$ and taking the $J_{a}^{i}(x)\to0$ limit, we obtain
\begin{equation}
\begin{split}
 &\int d^{d+1}x' \, \int {d^{d+1}x}\,\bar\epsilon_{\beta}(x') \delta^{\beta}_{R} \bar{\varphi}^b_A [D^{-1}]_{ba}^{AR}(x'-x)  \delta^{\alpha}_A \bar{\varphi}_R^a\epsilon_{\alpha}(x)\\
&\qquad =-\lim_{J\to0} \int d^{d+1}x'\,\delta_R\varphi_j^b(x')\frac{\delta }{\delta \varphi^b_j(x')} \langle\delta_AS\rangle_J,
 \end{split}
\label{eq:basicRelation}
\end{equation}
where we interchange the left- and right-hand sides.
The left-hand side represents the inverse of the retarded Green function in the NG mode channel. We would like to express the right-hand side in the language of symmetry.
Naively, one might think that the right-hand side in Eq.~\eqref{eq:basicRelation} has the form $-\langle \delta_{R}\delta_{A}S\rangle$.
This is not the case. In general, a transformation of fields and the expectation value are not commutative due to fluctuations, i.e., $\langle \delta_{R}\mathcal{O}\rangle_{J}\neq\delta_{R}\langle \mathcal{O}\rangle_{J}$,
where we define
\begin{align}
\delta_{R}\mathcal{O} &:= \int d^{d+1}x\, \bar{\epsilon}_\alpha(x)\frac{\delta\mathcal{O}}{\delta \chi^{a}_i(x)} \delta_R^\alpha\chi_i^a(x),\\
\delta_{R}\langle \mathcal{O}\rangle_{J} &:= 
  \int d^{d+1}x \, \delta_{R} \varphi^a_i(x)  \frac{\delta}{\delta  \varphi_{i}^a(x)}\langle{\mathcal{O}}\rangle_J.
\end{align}
Recall that $\chi_i^a(x)$ is the elementary field of theory and all local operators are polynomials of $\chi_i^a$.
The equality holds if $\mathcal{O}$ is a linear function of $\phi_{R}^{a}$ and $\phi_{A}^{a}$.
To see the explicit relation between $\langle \delta_{R}\mathcal{O}\rangle_{J}$ and $\delta_{R}\langle \mathcal{O}\rangle_{J}$,
let us consider the Ward--Takahashi identity for $\delta_{R}\mathcal{O}$, which leads to
\begin{align}
\langle \delta_{R} \mathcal{O} \rangle_J= -\ri\langle  \delta_{R} S\,\mathcal{O}  \rangle_J - \ri\int d^{d+1}x'\, J_a^{i}(x') \langle\delta_{R} \phi^a_i(x')\,  \mathcal{O}\rangle_J .
\label{eq:deltaRO0}
\end{align}
We note that since the action is not invariant under this transformation, $\ri\langle  \delta_{R} S\, \mathcal{O} \rangle_J$ does not vanish even when $\bar\epsilon_\beta(x)$ is constant. 
For the trivial operator $\mathcal{O}=1$,  Eq.~\eqref{eq:deltaRO0} reduces to 
\begin{equation}
\ri\langle \delta_{R} S \rangle_J + \ri\int d^{d+1}x\, J_a^{i}(x) \delta_{R} \varphi^a_i(x)  =0. 
\label{eq:deltaRS}
\end{equation}
Noting that $\langle\mathcal{O}\phi_{j}^{b}(x) \rangle_{J}=\delta \langle\mathcal{O}\rangle_{J}/\delta\ri J_{b}^{j}(x)+\langle\phi_{j}^{b}(x)\rangle_{J}\langle\mathcal{O}\rangle_{J}$, and 
from Eq.~\eqref{eq:deltaRS},
we can express $\langle\delta_{R} \mathcal{O}\rangle_J$ as
\begin{align}
\langle\delta_{R} \mathcal{O}\rangle_J
&= -\ri\langle{\delta_{R} S\,\mathcal{O}}\rangle_J-\ri\int d^{d+1}x'\,J^{i}_a(x') \bar\epsilon_{\alpha}(x')\ri[T^\alpha]^a_{~b}\varepsilon_i^{~j} \Bigl(\frac{\delta}{\delta \ri J^{j}_b(x')}\langle{\mathcal{O}}\rangle_J+\langle{\phi_{j}^b(x')}\rangle_J\langle{\mathcal{O}}\rangle_J  \Bigr)\notag\\
&= -\ri\langle{ \delta_{R} S\,\mathcal{O}}\rangle_{\text{c};J}
-\ri\int d^{d+1}x'\,J^{i}_a(x')\bar\epsilon_{\alpha}(x')\ri[T^\alpha]^a_{~b}\varepsilon_i^{~j} \frac{\delta}{\delta \ri J^{j}_b(x')}\langle{\mathcal{O}}\rangle_J. 
\label{eq:deltaRO}
\end{align}
The subscript c denotes the connected diagram defined as $\average{{ \mathcal{O}_1\mathcal{O}_2}}_{\text{c};J}:=\average{{\Delta \mathcal{O}_1\Delta \mathcal{O}_2}}_{J}$ with $\Delta \mathcal{O}:=\mathcal{O}-\average{{\mathcal{O}}}_J$.
We would like to eliminate the explicit dependence of $J^{j}_b(x)$ in Eq.~\eqref{eq:deltaRO}. 
For this purpose, consider the functional derivative of Eq.~\eqref{eq:deltaRS} with respect to $\varphi^b_j(x')$, which becomes
\begin{align}
\frac{\delta}{\delta\varphi_j^b(x')}\ri\langle \delta_{R} S \rangle_J 
+ \ri\int d^{d+1}x\, 
 [D^{-1}]_{ba}^{j i}(x',x;\varphi)\delta_{R} \varphi^a_i(x) 
+\ri  J_a^{i}(x')\bar\epsilon_{\alpha}(x')\ri[T^\alpha]^a_{~b}\varepsilon_i^{~j} =0.
\label{eq:deltaRS2}
\end{align}
Substituting this into Eq.~\eqref{eq:deltaRO}, we can express $\langle\delta_{R} \mathcal{O}\rangle_J$ as
\begin{align}
\langle\delta_{R} \mathcal{O}\rangle_J
&= -\ri\langle{\delta_{R} S\, \mathcal{O}}\rangle_{\text{c};J}
+
\int d^{d+1}x'\,\frac{\delta}{\delta\varphi_j^b(x')}\ri\langle \delta_{R} S \rangle_J \frac{\delta}{\delta \ri J^{j}_b(x')}\langle{\mathcal{O}}\rangle_J\notag\\
&\quad+ \ri\int d^{d+1}x'\,\int d^{d+1}x\,   [D^{-1}]_{ba}^{j i}(x',x;\varphi)\delta_{R} \varphi^a_i(x)  \frac{\delta}{\delta  \ri J^{j}_b(x')}\langle{\mathcal{O}}\rangle_J.
 \label{eq:deltaRO2}
\end{align}
In order to obtain a more compact expression, we introduce the projection operator defined as 
\begin{equation}
\begin{split}
\mathcal{P}_{\phi}\mathcal{O}&:= \int d^{d+1}x'\, \Delta\phi_j^b(x')\frac{\delta}{\delta\varphi_j^b(x')}\langle  \mathcal{O} \rangle_J\\
&=\ri\int d^{d+1}x'\, \int d^{d+1}x\,\Delta\phi_j^b(x')  [D^{-1}]^{ji}_{ba}(x',x;\varphi) \langle \phi_i^a(x) \mathcal{O}\rangle_{\text{c};J}.
\label{eq:Projphi}
\end{split}
\end{equation}
Here, we assumed that $\mathcal{O}$ does not explicitly depend on $\varphi$.
We also introduce $\mathcal{Q}_{\phi}:=1-\mathcal{P}_{\phi}$, which satisfies $\mathcal{P}_{\phi}^{2}=\mathcal{P}_{\phi}$, $\mathcal{Q}^{2}_{\phi}=\mathcal{Q}_{\phi}$, and $\mathcal{P}_{\phi}\mathcal{Q}_{\phi}=\mathcal{Q}_{\phi}\mathcal{P}_{\phi}=0$. By construction $\mathcal{Q}_{\phi}\Delta\phi_{i}^{a}(x)=0$, that is, the projection operator $\mathcal{Q}_{\phi}$ removes the linear component of $\Delta \phi_{i}^{a}(x)$ from the operator. 
The first two terms on the right-hand side of Eq.~\eqref{eq:deltaRO2} are simply expressed as
\begin{align}
-\ri\langle{ \delta_{R} S\,\mathcal{O}}\rangle_{\text{c};J}+
\int d^{d+1}x'\,\frac{\delta}{\delta\varphi_j^b(x')}\ri\langle \delta_{R} S \rangle_J \frac{\delta}{\delta \ri J^{j}_b(x')}\langle{\mathcal{O}}\rangle_J
= -\ri\langle{ (\mathcal{Q}_\phi\delta_{R} S)\, (\mathcal{Q}_\phi\mathcal{O}})\rangle_{\text{c};J}.
\end{align}
Noting that $  [D^{-1}]_{ba}^{j i}(x',x;\varphi)=  {\delta J_b^{j}(x')}/{\delta\varphi_i^a(x)}$, and using the chain rule,
we obtain the last term of Eq.~\eqref{eq:deltaRO2} as
\begin{align}
 \ri\int d^{d+1}x'\,\int d^{d+1}x\,[D^{-1}]_{ba}^{j i}(x',x;\varphi)\delta_{R} \varphi^a_i(x)  \frac{\delta}{\delta \ri J^{j}_b(x')}\langle{\mathcal{O}}\rangle_J
 =
  \int d^{d+1}x \,\delta_{R} \varphi^a_i(x)  \frac{\delta}{\delta  \varphi_{i}^a(x)}\langle{\mathcal{O}}\rangle_J.
\end{align}
Eventually, we arrive at the expression,
\begin{equation}
\begin{split}
\langle\delta_{R} \mathcal{O}\rangle_J=
\delta_{R}\langle{\mathcal{O}}\rangle_J
  -\ri\langle{ (\mathcal{Q}_\phi\delta_{R} S)\,( \mathcal{Q}_\phi\mathcal{O}})\rangle_{\text{c};J}.
\end{split}
\end{equation}
If we choose $\mathcal{O}= \delta_AS$ and take the $J\to0$ limit, Eq.~\eqref{eq:deltaRO} becomes
\begin{align}
\begin{split}
\lim_{J\to0}  \int d^{d+1}x'\, \delta_{R} \bar{\varphi}^b_j(x') \frac{\delta}{\delta  \varphi_{j}^b(x')}\langle{\delta_AS}\rangle_J
&=
\langle\delta_{R}\delta_AS\rangle
+\ri\langle{ (\mathcal{Q}_\phi\delta_{R} S)\, (\mathcal{Q}_\phi \delta_AS)}\rangle_\text{c}.
\end{split}
\label{eq:limit}
\end{align}
Substituting Eq.~\eqref{eq:limit} into Eq.~\eqref{eq:basicRelation}, we obtain
\begin{align}
\begin{split}
 \int d^{d+1}x'\, \int {d^{d+1}x}\,\bar\epsilon_{\beta}(x') \delta^{\beta}_{R} \bar{\varphi}^b_A [D^{-1}]_{ba}^{AR}(x'-x)  \delta^{\alpha}_A \bar{\varphi}_R^a\epsilon_{\alpha}(x)
&=
-\langle\delta_{R}\delta_AS\rangle
-\ri\langle{ (\mathcal{Q}_\phi\delta_{R} S)\, (\mathcal{Q}_\phi \delta_AS)}\rangle_\text{c}.
\label{eq:formula1}
\end{split}
\end{align}
This expression gives the relation between the inverse of the retarded Green function and expectation values of operators.
The indices $a$ and $b$ in $[D^{-1}]_{ba}^{AR}(x'-x)$ include not only NG fields but also other fields with gapped modes.
Therefore, our next step is to find the inverse of the retarded Green function consisting of only the NG fields.

\subsection{NG modes and their low-energy coefficients} \label{sec:dispersion}
The dispersion relations or the positions of poles can be obtained by solving $\det D^{-1} (k) =0$ in momentum space.
Probability conservation ensures that $[D^{-1}]_{ab}^{RR}=0$, so that $\det D^{-1} (k)=\det [D^{-1}]^{AR} (k) \det [D^{-1}]^{RA} (k)=0$. Here, we will find the dispersion relation for the retarded Green function. 
$[D^{-1}]_{ba}^{AR} (k)$ contains not only NG modes but also gapped modes, and they can mix.
Therefore, we need to carefully  analyze the inverse of the retarded Green function  $[D^{-1}]_{ba}^{AR}(k)$.
To separate the NG modes and gapped modes, we decompose fields $\phi_i^a(x)$ as 
$\phi_i^a(x)=\delta_A^\alpha\bar{\varphi}_R^a \pi_{i \alpha}(x) + M^{a\tilde{\alpha}}\Phi_{i \tilde{\alpha}}(x)$.
Here, the $\pi_{i \alpha}(x) $ represent NG fields, with the $\Phi_{i \tilde{\alpha}}(x)$ gapped fields\footnote{
We note this gapped fields do not correspond to the gapped or damping modes appeared in Eq.~\eqref{eq:gappedMode}. 
$\Phi_{i \tilde{\alpha}}(x)$ includes a ``mass'' term, i.e., $\det{[D^{-1}_{\Phi\Phi}}](0,\bm{0})\neq0$,
while $G_{\pi}^{-1}$ does not include it.  $G_{\pi}^{-1}(k)$ vanishes at $\omega=0$ and $\bm{k}=\bm{0}$.
}.
In this basis, the inverse of the retarded Green function is expressed as
\begin{align}
[D^{-1}]_{ba}^{AR} (k) &\to\begin{pmatrix}
[D^{-1}_{\pi\pi}]^{\beta\alpha} (k)   & [D^{-1}_{\pi\Phi}]^{\beta\tilde{\alpha}} (k)  \\
[D^{-1}_{\Phi\pi}]^{\tilde{\beta}\alpha}(k)   & [D^{-1}_{\Phi\Phi}]^{\tilde{\beta}\tilde{\alpha}} (k)
\end{pmatrix}\notag\\
&=
\begin{pmatrix}
\delta_R^\beta\bar{\varphi}_A^b \delta_A^\alpha\bar{\varphi}_R^a [D^{-1}]_{ba}^{AR} (k)   &\delta_R^\beta\bar{\varphi}_A^b M^{ a \tilde{\alpha}} [D^{-1}]_{ba}^{AR}(k)   \\
M^{ b \tilde{\beta} } \delta_A^\alpha\bar{\varphi}_R^a [D^{-1}]_{ba}^{AR} (k)  & M^{b \tilde{\beta}} M^{ a \tilde{\alpha}}[D^{-1}]_{ba}^{AR}  (k)
\end{pmatrix}.
\end{align}
The determinant of the inverse of the retarded Green function can be decomposed into 
$\det [D^{-1}]^{AR}(k)   =\det{[D^{-1}_{\Phi\Phi}}](k) \det [G_\pi^{-1}](k)$ with 
\begin{align}
[G_\pi^{-1}(k)]^{\beta\alpha}=\left[D^{-1}_{\pi\pi}(k)- D^{-1}_{\pi\Phi}(k)\frac{1}{D^{-1}_{\Phi\Phi}(k)} D^{-1}_{\Phi\pi}(k)\right]^{\beta\alpha}.
\label{eq:InverseDpi}
\end{align}
We are interested in the dispersion relation of the NG modes, which can be found as the solution of $\det G_\pi^{-1}(k)=0$.
Our purpose now is  to express $[{G}_\pi^{-1}(k)]^{\beta\alpha}$ by using the correlation functions of currents.
For this purpose, we rewrite Eqs.~\eqref{eq:basic} and \eqref{eq:deltaRS2} at the limit $J_{a}^{i}(x)\to 0$.
Noting that
\begin{align}
\lim_{J\to0} \frac{\delta }{\delta \varphi_j^{b}(x')} \average{\mathcal{O}}_J
=  \int d^{d+1}x\, \frac{\delta  \ri J^i_{a}(x)}{\delta \varphi_j^{b}(x')}\frac{\delta}{\delta \ri J^i_{a}(x)} \average{\mathcal{O}}
 =  \ri\int d^{d+1}x\, [D^{-1}]^{ji}_{ba}(x'-x) \average{ \phi_i^{a}(x) \mathcal{O}}_{\text{c}},
\end{align}
we  have the following equations:
\begin{align}
  \ri\int {d^{d+1}x}\,  [D^{-1}]_{ba}^{jR}(x'-x) \epsilon_\alpha(x)  \delta^{\alpha}_A \bar{\varphi}_R^{a}&=
  \int d^{d+1}x \,[D^{-1}]^{ji}_{ba}(x'-x) \average{  \phi_i^{a}(x) \delta_A S}_{\text{c}},
  \label{eq:delphiA}\\
\ri \int d^{d+1}x\, 
\bar\epsilon_{\alpha}(x)\delta_R^{\alpha} \bar{\varphi}_A^{a} [D^{-1}]^{A j}_{ab}(x-x')&=
\int d^{d+1}x\,\average{  \delta_RS  \phi_i^{a}(x)}_{\text{c}}  [D^{-1}]^{ij}_{ab}(x-x'). 
\label{eq:delphiB}
\end{align}
To avoid complicated indices, we write Eqs.~\eqref{eq:delphiA} and \eqref{eq:delphiB} in  matrix form:
\begin{align}
\ri D_{\Phi\pi}^{-1}\epsilon&=D^{-1}_{\Phi \pi} S_\pi
+D^{-1}_{\Phi \Phi}S_\Phi, \label{eq:Spi1}\\
\ri \bar\epsilon D_{\pi\Phi}^{-1}&=
\bar{S}_\pi D_{\pi\Phi}^{-1}
+\bar{S}_\Phi D_{\Phi\Phi}^{-1}, \label{eq:Spi2}
\end{align}
where we define $[S_\pi]_{j\alpha}(x) := \average{\pi_{j\alpha}(x)\delta_A S}$,  $[S_\Phi]_{i\tilde{\alpha}}(x) := \average{\Phi_{i\tilde{\alpha}}(x)\delta_A S}$,
$[\bar{S}_\pi]_{j\alpha}(x) := \average{\delta_R S\pi_{j\alpha}(x)}$,  and $[\bar{S}_\Phi]_{i\tilde{\alpha}}(x) := \average{\delta_R S\Phi_{i\tilde{\alpha}}(x)}$.
Here, we introduce a projection operator $\mathcal{P}_\pi$ and $\mathcal{Q}_\pi=1-\mathcal{P}_\pi$ defined as
\begin{equation}
\begin{split}
\mathcal{P}_\pi \mathcal{O} :=   \ri\int d^{d+1}x'\, \int d^{d+1}x\,  \Delta\pi_{j\beta}(x') [G_{\pi}^{-1}]^{ji;\beta\alpha}(x'-x)\average{ \pi_{i\alpha}(x) \mathcal{O}}_{\text{c}}.
\label{eq:Ppi}
\end{split}
\end{equation}
Unlike $\mathcal{P}_{\phi}$ defined in Eq.~\eqref{eq:Projphi}, $\mathcal{P}_\pi$ projects to only the NG fields.
This is natural since we would like to focus on only the NG modes.
By the definition of $\mathcal{Q}_{\phi}$, $-\ri\average{{ (\mathcal{Q}_\phi\delta_R S)\, (\mathcal{Q}_\phi\delta_A S)}}_{\text{c}}$ 
is expanded as
\begin{align}
-\ri\average{{ (\mathcal{Q}_\phi\delta_R S)\, (\mathcal{Q}_\phi\delta_A S)}}_{\text{c}}
&=-\ri\average{{ \delta_R S \delta_A S}}_{\text{c}} 
- \bar S_\pi D_{\pi\pi}^{-1}S_\pi
-\bar S_\Phi D_{\Phi\pi}^{-1}S_\pi
-\bar S_\pi D_{\pi\Phi}^{-1}S_\Phi
-\bar S_\Phi D_{\Phi\Phi}^{-1}S_\Phi.
\end{align}
Similarly, for $\mathcal{Q}_\pi$  we find that
\begin{equation}
\begin{split}
-\ri\average{{ (\mathcal{Q}_\pi\delta_R S)\, (\mathcal{Q}_\pi\delta_A S)}}_{\text{c}}&=
-\ri\average{{ \delta_R S \delta_A S}}_{\text{c}} 
- \bar S_\pi G_{\pi}^{-1}S_\pi.
\end{split}
\end{equation}
The difference of these is, from Eq.~\eqref{eq:InverseDpi},
\begin{equation}
\begin{split}
&\ri\average{{ (\mathcal{Q}_\pi\delta_R S)\, (\mathcal{Q}_\pi\delta_A S)}}_{\text{c}} 
-\ri\average{{ (\mathcal{Q}_\phi\delta_R S)\, (\mathcal{Q}_\phi\delta_A S)}}_{\text{c}} \\
&=
\bar S_\pi\left[D^{-1}_{\pi\pi}- D^{-1}_{\pi\Phi}\frac{1}{D^{-1}_{\Phi\Phi}} D^{-1}_{\Phi\pi}\right] S_\pi
- \bar S_\pi D_{\pi\pi}^{-1}S_\pi
-\bar S_\Phi D_{\Phi\pi}^{-1}S_\pi
-\bar S_\pi D_{\pi\Phi}^{-1} S_\Phi
-\bar S_\Phi D_{\Phi\Phi}^{-1}S_\Phi\\
&=
-(S_\pi D^{-1}_{\pi\Phi}
+S_\Phi D^{-1}_{\Phi\Phi})
\frac{1}{D^{-1}_{\Phi\Phi}} 
(D^{-1}_{\Phi\pi}
S_\pi
+D^{-1}_{\Phi\Phi}S_\Phi).
\label{eq:difference}
\end{split}
\end{equation}
Substituting Eqs.~\eqref{eq:Spi1} and \eqref{eq:Spi2} into Eq.~\eqref{eq:difference}, we can write the difference as 
\begin{align}
\ri\average{{ (\mathcal{Q}_\pi\delta_R S)\, (\mathcal{Q}_\pi\delta_A S)}}_{\text{c}} -\ri\average{{ (\mathcal{Q}_\phi\delta_R S)\, (\mathcal{Q}_\phi\delta_A S)}}_{\text{c}}
&=
\bar{\epsilon} D_{\pi\Phi}^{-1}\frac{1}{D_{\Phi\Phi}^{-1}}D_{\Phi\pi}^{-1}\epsilon.
\label{eq:differeince2}
\end{align}
Recall Eq.~\eqref{eq:formula1}, which is  written in the compact notation as
\begin{equation}
\begin{split}
\bar{\epsilon} D^{-1}_{\pi\pi} \epsilon =- \langle\delta_{R}\delta_{A}S \rangle -\ri \langle(\mathcal{Q}_{\phi}\delta_{R}S)\,(\mathcal{Q}_{\phi}\delta_{A}S)\rangle_\text{c}.
\label{eq:compact}
\end{split}
\end{equation}
Substituting Eqs.~\eqref{eq:differeince2} and \eqref{eq:compact} into Eq.~\eqref{eq:InverseDpi}, we finally obtain
\begin{equation}
\begin{split}
\bar{\epsilon} G_{\pi}^{-1} \epsilon = 
-\langle\delta_{R}\delta_AS\rangle
-\ri\langle{ (\mathcal{Q}_\pi\delta_{R} S)\, (\mathcal{Q}_\pi \delta_AS)}\rangle_\text{c}.
\label{eq:NonperturbativeRelation}
\end{split}
\end{equation}
This equation gives the relation between the inverse of the retarded Green function for NG fields and the expectation value of operators.

Let us evaluate the right-hand side of Eq.~\eqref{eq:NonperturbativeRelation}.
The first term is expressed as
\begin{equation}
\begin{split}
\langle\delta_{R}\delta_AS\rangle&=
-\int d^{d+1}x\,\langle\delta_{R}j_{A}^{\alpha\mu}(x)\rangle\partial_{\mu}\epsilon_{\alpha}(x).
\label{eq:deltaRAS}
\end{split}
\end{equation}
The local transformation of the current can be expanded as
\begin{equation*}
\delta_{R}j_{A}^{\alpha\mu}(x) = 
\bar{\epsilon}_{\beta}(x)\delta_{R}^{\beta}j_{A}^{\alpha\mu}(x)+
\partial_{\nu}\bar{\epsilon}_{\beta}(x) \mathcal{S}^{\beta\alpha;\nu\mu}(x)+\cdots.
\tag{\ref{eq:deltaRJA}}
\end{equation*}
The first term $\delta_{R}^{\beta}j_{A}^{\alpha\mu}(x)$ is the transformation of $j_{A}^{\alpha\mu}(x)$ under  $\delta^{\alpha}_{R}$.
$\mathcal{S}^{\beta\alpha;\mu}(x)$ is analogus to the Schwinger term in quantum field theory.
Similarly,  $\delta_{R} S $ is expanded as
\begin{equation*}
\delta_{R} S = \int d^{d+1}x\, \bigl[\bar{\epsilon}_{\alpha}(x) \brokenPotential_{R}^{\alpha}(x) 
-\partial_{\mu}\bar{\epsilon}_{\alpha}(x) j_{R}^{\alpha\mu}(x) 
\bigr].
\tag{\ref{eq:deltaRS3}}
\end{equation*}
Substituting Eqs.~\eqref{eq:deltaRS3}, \eqref{eq:deltaRJA} and \eqref{eq:deltaRAS} into Eq.~\eqref{eq:NonperturbativeRelation},
we find the inverse of the retarded Green function in momentum space as
\begin{equation}
\begin{split}
[G_\pi^{-1}(k)]^{\beta\alpha}  = -\ri k_{\mu}\langle \delta_{R}^{\beta}j_{A}^{\alpha}(0)\rangle
+k_{\nu} k_{\mu}\langle S^{\beta\alpha;\nu\mu}(0)\rangle
-G_{{h}_{R}j_{A}}^{\beta\alpha;\mu}(k)  \ri k_{\mu}
-G_{{j}_{R}j_{A}}^{\beta\alpha;\nu\mu}(k)  k_{\nu}k_{\mu},
\cdots,
\label{eq:Gk}
\end{split}
\end{equation}
with
\begin{align}
G_{h_{R}j_{A}}^{\beta\alpha;\mu}(k)  &:=\ri\int d^{d+1}x\,e^{\ri k\cdot x}\langle \bigl(\mathcal{Q}_{\pi}h_{R}^{\beta}(x)\bigr)\,\bigl(\mathcal{Q}_{\pi}j_{A}^{\alpha\mu}(0)\bigr)\rangle_\text{c}, \\
G_{j_{R}j_{A}}^{\beta\alpha;\nu\mu}(k)  &:=\ri\int d^{d+1}x\,e^{\ri k\cdot x}\langle \bigl(\mathcal{Q}_{\pi}j_{R}^{\beta\nu}(x)\bigr)\,\bigl(\mathcal{Q}_{\pi}j_{A}^{\alpha\mu}(0)\bigr)\rangle_\text{c}.
\label{eq:DjRjA}
\end{align}
Here, $\cdots$ denotes the contribution coming from the expansion of $\delta_{R}^{\beta}j_{A}^{\alpha\mu}(x)$, which is local and can be directly evaluated from Eq.~\eqref{eq:deltaRJA}. We are interested in the low-energy behavior, so that we expand the inverse of the retarded Green function in terms of $k^{\mu}$:
\begin{align}
\begin{split}
[G_\pi^{-1}(k)]^{\beta\alpha} = C^{\beta\alpha}  -\ri C^{\beta\alpha;\mu} k_\mu +C^{\beta\alpha;\nu \mu}k_\nu k_\mu +\cdots,
\end{split}
\end{align}
and the first three coefficients are then
\begin{align}
C^{\beta\alpha} &= 0, \label{eq:C}\\
  C^{\beta\alpha;\mu}
&=
\average{\delta_{R}^{\beta} j_{A}^{\alpha\mu}(0)}
+\lim_{k\to0}G_{h_{R}j_{A}}^{\beta\alpha;\mu}(k), 
\label{eq:Cmu}\\
  C^{\beta\alpha;\nu \mu}
&=\average{ \mathcal{S}^{\beta\alpha;\nu\mu}(0)}
-\lim_{k\to0}G_{j_{R}j_{A}}^{\beta\alpha;\nu\mu}(k) 
-\ri \lim_{k\to0}\frac{\partial}{\partial k_\nu}G_{h_{R}j_{A}}^{\beta\alpha;\mu}(k),
\label{eq:C10formula2}
\end{align}
which corresponds to  Eqs.~\eqref{eq:invG}-\eqref{eq:C10formula}.
These are the formulae that we would like to derive in this paper.
The important assumption of these formulae is that  Eqs.~\eqref{eq:Cmu} and \eqref{eq:C10formula2} are nondivergent in the limit $k^{\mu}\to0$.
If this is not the case, we need to keep the momentum dependence, which may change the power of the momentum.
The reality condition in Eq.~\eqref{eq:RealityCondition} implies that all the $C^{\alpha\cdots;\nu \mu}$ of the derivative expansion are real.
If the system is isolated, $\average{\delta_{R}^{\beta} j_{A}^{\alpha\mu}(0)}$ vanishes because the action is invariant under $\delta_{R}$.
In this case, $C^{\beta\alpha;0}$ reduces to $\average{\delta_{R}^{\beta} j_{A}^{\alpha0}(0)}$, which Watanabe and Murayama obtained in the effective Lagrangian approach at zero temperature~\cite{Watanabe:2012hr}. 
We note that $C^{\beta\alpha;i}$ vanishes in isolated systems because of the stability of the system. If this is not the case, one can find a more stable solution, where the translational symmetry is spontaneously broken.

When the symmetry is not exact but approximate, the action is not invariant under the $\delta_{A}$ transformation.
Under the local transformation, $\delta_{A}S$ has the form,
\begin{equation}
\begin{split}
\delta_{A}S = 
\int d^{d+1}x\,[\epsilon_{\alpha}(x)h^{\alpha}_{A}(x)-j_{A}^{\alpha\mu}(x)\partial_{\mu}\epsilon_{\alpha}(x) ],
\end{split}
\end{equation}
where $h^{\alpha}_{A}(x)$ represents the explicit breaking term, which  leads to the coefficient in Eq.~\eqref{eq:C} as
\begin{equation}
\begin{split}
C^{\beta\alpha} = -\langle \delta_{R}^{\beta}h_{A}^{\alpha}(x)\rangle.
\end{split}
\end{equation}
This term gives the Gell-Man--Oaks--Renner relation in open systems~\cite{GellMann:1968rz}.
The explicit breaking term $h^{\alpha}_{A}(x)$ also modifies $C^{\beta\alpha;\mu}$ and $C^{\beta\alpha;\nu\mu}$, which can be straightforwardly evaluated.

\subsection{Example}
Let us see how our formalism works in a simple model with $SU(2)\times U(1)$ symmetry.
This model is known to exhibit both type-A and type-B NG modes in an isolated system at finite density~\cite{Miransky:2001tw,Schafer:2001bq}.
The classical version of this model in an open system is employed for the analysis of NG modes~\cite{Minami:2018oxl}.
The action that we consider has the form, $S=S_{1}-S_{2}+S_{12}$ with
\begin{align}
S_{i} &=   \int {d^{d+1}x}\, \Bigl[ (\partial_{t}-\ri \mu)\varphi_{i}^{\dag}(\partial_{t}+\ri \mu)\varphi_{i} - \bm{\nabla}\varphi_{i}^{\dag}\bm{\nabla}\varphi_{i}
 -\lambda (\varphi_{i}^{\dag}\varphi_{i})^{2}],\\
 S_{12}&= \int {d^{d+1}x}\, [  -\gamma\varphi_1^{\dag}\partial_{t}\varphi_2  +\gamma\varphi_2^{\dag}\partial_{t}\varphi_1
+\ri \kappa(\varphi_{1}^{\dag}-\varphi_{2}^{\dag})(\varphi_{1}-\varphi_{2})]
\end{align}
in the $1/2$ basis. Here, $\varphi_{i}=(\varphi^1_{i},\varphi^2_{i})$  is the two-component complex scalar field.
The coefficients $\mu$, $\lambda$, $\gamma$, and $\kappa$ are the chemical potential, the coupling constant, 
the friction and the fluctuation coefficient, respectively. The dagger $\dag$ represents the Hermitian conjugate. 
If $\gamma$ and $\kappa$ vanish, this system reduces to   the isolated one.
It is useful to express the action in the Keldysh basis, where $\varphi_{R}:=(\varphi_{1}+\varphi_{2})/2$ and $\varphi_{A}:=\varphi_{1}-\varphi_{2}$, as 
\begin{equation}
\begin{split}
 S&= \int {d^{d+1}x}\, \Bigl[ (\partial_{t}-\ri \mu)\varphi_{R}^{\dag}(\partial_{t}+\ri \mu)\varphi_{A}+(\partial_{t}-\ri \mu)\varphi_{A}^{\dag}(\partial_{t}+\ri \mu)\varphi_{R}
  - \bm{\nabla}\varphi_{R}^{\dag}\bm{\nabla}\varphi_{A}- \bm{\nabla}\varphi_{A}^{\dag}\bm{\nabla}\varphi_{R}\\
&\qquad\qquad\quad  -2\lambda\Bigl(|\varphi_{R}|^{2}+\frac{|\varphi_{A}|^{2}}{4}\Bigr)(\varphi_{R}^{\dag}\varphi_{A} +\varphi_{A}^{\dag}\varphi_{R})
  +\gamma\varphi_R^{\dag}\partial_{t}\varphi_{A}-\gamma\varphi_{A}^{\dag}\partial_{t}\varphi_R
+\ri \kappa\varphi_{A}^{\dag}\varphi_{A}
\Bigr].
\end{split}
\end{equation}
This action is invariant under an $SU(2)\times U(1)$ transformation, 
$\varphi_i \to \varphi_i+ \ri \epsilon_\alpha T^\alpha \varphi_i$,
where $T^0$ is the $U(1)$ generator, and $T^a$ ($a=1,2,3$) are the $SU(2)$ generators satisfying the Lie algebra, $[T^\alpha,T^\beta]= \ri f^{\alpha\beta\gamma} T^\gamma$, where $f^{\alpha\beta\gamma}$ is the structure constant with $f^{0\beta\gamma}=f^{\alpha0\gamma}=f^{\alpha\beta0}=0$, and $f^{abc}=\epsilon^{abc}$ (the indices $a,b$, and $c$ run from $1$ to $3$). $\epsilon^{abc}$ is the Levi--Civita symbol.
We choose the normalization of the generators as $\mathrm{tr} \, T^\alpha T^\beta = \delta^{\alpha\beta}/2$.
From Eq.~\eqref{eq:deltaAS}, the Noether current is given as
\begin{align}
j_{A}^{\alpha 0} &= -(\partial_{t}-\ri\mu-\gamma)\varphi_A^\dag  \ri T^{\alpha}\varphi_{R} + \varphi_{A}^{\dag}\ri T^{\alpha}(\partial_{t}+\ri \mu)\varphi_R\notag\\
&\quad +  \varphi_{R}^{\dag}\ri T^{\alpha}(\partial_{t}+\ri\mu-\gamma)\varphi_A  -(\partial_{t}-\ri \mu)\varphi_R^\dag  \ri T^{\alpha}\varphi_{A},\\
j_{A}^{\alpha i} &= \bm{\nabla} \varphi_A^\dag  \ri T^{\alpha}\varphi_{R} - \varphi_{A}^{\dag}\ri T^{\alpha}\bm{\nabla}\varphi_R
 + \bm{\nabla}\varphi_R^\dag  \ri T^{\alpha}\varphi_{A}-  \varphi_{R}^{\dag}\ri T^{\alpha}\bm{\nabla}\varphi_A  .
\end{align}
Let us solve this model to find the dispersion relation of NG modes within the mean field approximation~\cite{Sieberer:2015svu}. 
The stationary solution is obtained from 
\begin{equation}
\begin{split}
\frac{\delta S}{\delta \varphi_{A}}&=\frac{\delta S}{\delta \varphi_{A}^{\dag}}=\frac{\delta S}{\delta \varphi_{R}}=\frac{\delta S}{\delta \varphi_{R}^{\dag}}=0,
\end{split}
\end{equation}
which leads to $|\varphi_{R}|^{2}=\mu^{2}/(2\lambda)$ and $\varphi_{A}=0$.
We parametrize the solution as $\varphi_{R}=\varphi_{0}:=(0,v/\sqrt{2})$ with $v=\sqrt{\mu^{2}/\lambda}$. Note that $\varphi_{R}=0$ is also a solution;  it is, however, unstable against small perturbations.
The order parameter is not invariant under  $(T^{0}-T^{3})\varphi_{0}\neq0$, $T^{1}\varphi_{0}\neq0$,$T^{2}\varphi_{0}\neq 0$, while $(T^{0}+T^{3})\varphi_{0}=0$;
thus the symmetry breaking pattern is $SU(2)\times U(1)\to U(1)_{T^{0}+T^{3}}$, whose broken generators are $T^{1}$, $T^{2}$ and $(T^{0}-T^{3})/2:= T^{3'}$.

In order to see the low-energy behavior,  we expand the field around the stationary solution as\footnote{We use slightly different normalization from that in Ref.~\cite{Minami:2018oxl}.}
\begin{equation}
\varphi_R=\frac{1}{\sqrt{2}}(\chi_R^{1}+\ri\chi_R^2, v+\psi_R^{1}+\ri\psi^{2}_R), \qquad
\varphi_A=\frac{1}{\sqrt{2}}(\chi_A^{1}+\ri\chi_A^2, \psi_A^{1}+\ri\psi^{2}_A).
\end{equation}
Then the action reduces to 
\begin{equation}
\begin{split}
S&=\int d^{d+1}x\,\Bigl[
\chi_{A}^{a}(-\partial_{t}^{2}-\gamma\partial_{t}+\bm{\nabla}^{2})\chi^{a}_{R}
+\psi_{A}^{a}(-\partial_{t}^{2}-\gamma\partial_{t}+\bm{\nabla}^{2})\psi_{R}^{a}+ \frac{\ri \kappa}{2}\bigl((\psi^{a}_{A})^{2}+(\chi^{a}_{A})^{2}\bigr)\\
&\qquad\qquad\quad-2\mu^{2}\psi^{1}_{A}\psi_{R}^{1}+2\mu\epsilon_{ab}(\chi_{A}^{a}\partial_{t}\chi_{R}^{b}+\psi_{A}^{a}\partial_{t}\psi_{R}^{b})
\Bigr]+\cdots .
\end{split}
\end{equation}
Here, the indices $a,b$ run from $1$ to $2$. We have dropped constant  and total derivative terms. ``$\cdots$'' denotes the nonlinear terms of the fields.
$\epsilon_{ab}$ is the antisymmetric tensor with $\epsilon_{12}=1$.
Since $\psi^{1}$  has the mass gap, we can get rid of $\psi^{1}$ to obtain the effective Lagrangian for gapless modes 
by using the equations of motion:
\begin{equation}
\begin{split}
(-\partial_{t}^{2}-\gamma\partial_{t}+\bm{\nabla}^{2}-2\mu^{2})\psi_{R}^{1}+2\mu\partial_{t}\psi_{R}^{2} +\ri \kappa\psi_{A}^{1} &= 0,\\
(-\partial_{t}^{2}+\gamma\partial_{t}+\bm{\nabla}^{2}-2\mu^{2})\psi_{A}^{1}+2\mu\partial_{t}\psi_{A}^{2}  &= 0.
\end{split}
\end{equation}
At the leading order of the derivative expansion, the solutions are 
\begin{equation}
\psi_{R}^{1}=\frac{1}{\mu}\partial_{t}\psi_{R}^{2} +\frac{\ri \kappa}{2\mu^{2}}\psi_{A}^{1},\qquad
\psi_{A}^{1}=\frac{1}{\mu}\partial_{t}\psi_{A}^{2}.
\end{equation}
Substituting the solutions into the action, we find
\begin{equation}
\begin{split}
S_\text{eff}&=\int d^{d+1}x\,\Bigl[2\mu\epsilon_{ab}\chi_{A}^{a}\partial_{t}\chi_{R}^{b}
+\chi_{A}^{a}(-\partial_{t}^{2}-\gamma\partial_{t}+\bm{\nabla}^{2})\chi^{a}_{R}
+\psi_{A}^{2}(-3\partial_{t}^{2}-\gamma\partial_{t}+\bm{\nabla}^{2})\psi_{R}^{2}\\
&\qquad\qquad\quad  + \frac{\ri \kappa}{2}\Bigl((\chi^{a}_{A})^{2}+(\psi^{2}_{A})^{2}+\frac{1}{\mu^{2}}(\partial_{t}\psi^{2}_{A})^{2}\Bigr)
\Bigr]+\cdots .
\label{eq:Seff}
\end{split}
\end{equation}
From this effective action, we can read the inverse of the retarded Green function:
\begin{equation}
\begin{split}
G^{-1}_{\pi} = \begin{pmatrix}
-\omega^{2}+\bm{k}^{2}-\ri \omega \gamma& 2\ri \mu \omega  & 0 \\
-2\ri \mu  \omega & -\omega^{2}+\bm{k}^{2}-\ri \omega \gamma&  0  \\
0 & 0 & -3\omega^{2}+\bm{k}^{2}-\ri \omega \gamma
\end{pmatrix}
\end{split}.
\end{equation}
Here, the first, second and third components correspond to $\chi^{1}$, $\chi^{2}$ and $\psi^{2}$, respectively.
The dispersion relations are obtained from $\det G_{\pi}^{-1}=0$ as
\begin{equation}
\begin{split}
\omega &= -\frac{\ri}{\gamma}|\bm{k}|^{2},\qquad
\omega =  \frac{|\bm{k}|^2}{4\mu^2+\gamma^2}(\pm2\mu- \ri \gamma)
\end{split}
\end{equation}
at small $\bm{k}$.
Therefore this system has one diffusion and one propagation modes\footnote{More precisely, there are one damping and one gapped propagation mode in addition to NG modes. The damping and the diffusion NG modes become one propagation mode at large $k$, which is called the gapped momentum state~\cite{Baggioli:2019jcm}.}.

Let us reproduce this result by using the matching formulae of Eqs.~\eqref{eq:invG}-\eqref{eq:C10formula}.
For this purpose, we need the explicit expression of $\delta_{R}^{\beta}j_{A}^{\mu}$, $\mathcal{S}^{\beta\alpha;\nu\mu}$, $j^{\beta\mu}_{R}$ and $h^{\beta}_{R}$.
Under the local transformation $\delta_{R} \varphi_{A}(x) =\bar{\epsilon}_{\alpha}(x)\ri T^{\alpha}\varphi_{R}(x)$ and 
$ \delta_{R} \varphi_{R}(x) = \bar{\epsilon}_{\alpha}(x)\ri T^{\alpha}\varphi_{A}(x)/4$,
$j_{A}^{\alpha\mu}(x)$ transforms as Eq.~\eqref{eq:deltaRJA}
with
\begin{align}
\delta_{R}^{\beta}j_{A}^{\alpha0}&= -f^{\beta\alpha\gamma}\Bigl((\partial_{t}-\ri \mu)\varphi_R^\dag   \ri T^{\gamma} \varphi_{R}
-  \varphi_{R}^{\dag} \ri T^{\gamma}(\partial_{t}+\ri\mu) \varphi_R 
\notag\\
 & +\frac{1}{4}(\partial_{t}-\ri\mu)\varphi_A^\dag  \ri T^{\gamma}\varphi_{A}
  -  \frac{1}{4}\varphi_{A}^{\dag} \ri T^{\gamma}(\partial_{t}+\ri \mu)\varphi_A\Bigr)\notag\\
&+\gamma\varphi_{R}^{\dag} \{T^{\beta},T^{\alpha}\} \varphi_R    -\frac{1}{4}\gamma\varphi_A^\dag   \{T^{\beta}, T^{\alpha}\}\varphi_{A},
 \\
\delta_{R}^{\beta}j_{A}^{\alpha i}&=f^{\beta\alpha\gamma}\Bigl(
  \bm{\nabla}\varphi_R^\dag   \ri T^{\gamma}\varphi_{R}
 -\varphi_{R}^{\dag} \ri T^{\gamma}\bm{\nabla}\varphi_R 
+\frac{1}{4} \bm{\nabla} \varphi_A^\dag   \ri T^{\gamma}\varphi_{A}
  -\frac{1}{4} \varphi_{A}^{\dag} \ri T^{\gamma}\bm{\nabla}\varphi_A
  \Bigr),\\
\mathcal{S}^{\beta\alpha;\nu\mu} &= -\eta^{\nu\mu}\Bigl(\varphi_{R}^{\dag} \{T^{\beta}, T^{\alpha}\}\varphi_R 
     +\frac{1}{4} \varphi_{A}^{\dag} \{T^{\beta} ,T^{\alpha}\}\varphi_A\Bigr),
\end{align}
where $\{T^{\beta},T^{\alpha}\}=T^{\beta}T^{\alpha}+T^{\alpha}T^{\beta}$.
The expectation values are evaluated within the mean-field approximation as
\begin{equation}
\begin{split}
\langle \delta_{R}^{\beta}j_{A}^{\alpha0} \rangle=\frac{v^{2}}{4}
\begin{pmatrix}
\gamma & 2\mu & 0\\
-2\mu & \gamma& 0\\
0 & 0 & \gamma
\end{pmatrix},
\qquad
\langle \mathcal{S}^{\beta\alpha;\nu\mu}  \rangle=
- \eta^{\nu\mu}\frac{v^{2}}{4}
\begin{pmatrix}
1 & 0 & 0\\
0 & 1 & 0\\
0 & 0 & 1
\end{pmatrix}.
\end{split}
\end{equation}
Here, the indices of the matrix are $1$, $2$, and $3'$, which are the indices of the broken generators.
The transformation of the action under local $\delta_{R}$ is evaluated as Eq.~\eqref{eq:deltaRS3}
with 
\begin{align}
\brokenPotential_{R}^{\beta} &=2\gamma\varphi_{R}^{\dag}\ri T^{\beta}\partial_{t}\varphi_R-\frac{1}{2}\gamma\varphi_A^{\dag}\ri T^{\beta}\partial_{t}\varphi_{A}
+\ri \kappa(\varphi_{A}^{\dag}\ri T^{\beta}\varphi_{R}-\varphi_{R}^{\dag}\ri T^{\beta}\varphi_{A}),
\\
j_{R}^{\beta0} &=
\varphi_{R}^{\dag}\ri T^{\beta}(\partial_{t}+\ri \mu)\varphi_{R}
-(\partial_{t}-\ri \mu)\varphi_{R}^{\dag} \ri T^{\beta}\varphi_{R}\notag\\
&\quad+\frac{1}{4}\varphi_{A}^{\dag}\ri T^{\beta}(\partial_{t}+\ri \mu)\varphi_{A}
-\frac{1}{4}(\partial_{t}-\ri \mu)\varphi_{A}^{\dag}\ri T^{\beta}\varphi_{A}
-\gamma\varphi_R^{\dag}\ri T^{\beta}\varphi_{R}+\frac{1}{4}\gamma\varphi_{A}^{\dag}\ri T^{\beta}\varphi_A,\\
j_{R}^{\beta i} &=\bm{\nabla}\varphi_{R}^{\dag}\ri T^{\beta}\varphi_{R}-\varphi_{R}^{\dag}\ri T^{\beta}\bm{\nabla}\varphi_{R}
+ \frac{1}{4}\bm{\nabla}\varphi_{A}^{\dag}\ri T^{\beta}\varphi_{A}- \frac{1}{4}\varphi_{A}^{\dag}\ri T^{\beta}\bm{\nabla}\varphi_{A}.
\end{align}

The projection operator $\mathcal{Q}_\pi$ removes the NG fields from operators, so that we obtain
\begin{align}
 \mathcal{Q}_\pi j_{A}^{3' 0} &= -\mu v \psi_{A}^{1}+\cdots,\\
  \mathcal{Q}_\pi j_{R}^{3'0}&=-\frac{1}{2}(2\mu+\ri\gamma)v\psi_{R}^{1}+\cdots,\\
 \mathcal{Q}_\pi \brokenPotential_{R}^{3'} &= \frac{\ri\gamma v}{2}\partial_{t}\psi_{R}^{1}+\cdots,
\end{align}
where the ``$\cdots$'' denote higher-order derivative terms and nonlinear terms of fields.
Other components have no linear term, so do not contribute to the correlation functions in the mean field approximation.
The correlation functions that contribute to $C^{\beta\alpha;\nu\mu}$ are 
\begin{equation}
\begin{split}
-\ri\int d^{d+1}x\, \average{    \bigl(\mathcal{Q}_\pi j^{{3'} \nu}_R(x)\bigr)\, \bigl(\mathcal{Q}_\pi j^{{3'}\mu}_A(0) \bigr)}_{\text{c}}
&=-\ri\delta^{\mu}_{0}\delta^{\nu}_{0}\frac{1}{2}(2\mu+\ri \gamma)\mu v^{2} \int d^{d+1}x\, \average{ \psi_{R}^{1}(x)\psi_{A}^{1}(0) }_{\text{c}}
\\
&=-\ri\delta^{\mu}_{0}\delta^{\nu}_{0}\frac{1}{2}(2\mu+\ri \gamma)\mu v^{2}  \lim_{k\to0} \frac{\ri}{k^{2}-2\mu^{2}+\ri\gamma k^{0}}\\
&=-\delta^{\mu}_{0}\delta^{\nu}_{0}\frac{ v^{2} }{4\mu}(2\mu+\ri \gamma),
\end{split}
\end{equation}
and
\begin{equation}
\begin{split}
 \lim_{k\to0}\frac{\partial}{\partial k_\nu}\int d^{d+1}x\,e^{\ri k\cdot x} \langle{\bigl( \mathcal{Q}_\pi\brokenPotential_{R}^{3'}(x)\bigr)\, \bigl(\mathcal{Q}_\pi j_A^{3'\mu}(0)\bigr)}\rangle_\text{c}&=
-\delta_{0}^{\mu}\frac{\ri\gamma\mu v^{2}}{2}  \lim_{k\to0}\frac{\partial}{\partial k_\nu}\int d^{d+1}x\,e^{\ri k\cdot x} \langle{\partial_{t}\psi^{1}_{R}(x)\psi^{1}_{A}(0)}\rangle_\text{c}\\
&=\delta_{0}^{\mu}\frac{\ri\gamma\mu v^{2}}{2}  \lim_{k\to0}\frac{\partial}{\partial k_\nu}\ri \omega\frac{\ri}{k^{2}-2\mu^{2}+\ri \gamma\omega}\\
&=\delta_{0}^{\mu}\delta_{0}^{\nu}\frac{ v^{2}}{4\mu} \ri\gamma.
\end{split}
\end{equation}
Collecting these results, we eventually find the low-energy coefficients as
\begin{align}
  C^{\beta\alpha;\mu}
= \delta^{\mu}_{0}\frac{v^{2}}{4}
\begin{pmatrix}
\gamma & 2\mu & 0\\
-2\mu & \gamma& 0\\
0 & 0 & \gamma
\end{pmatrix},\qquad
  C^{\beta\alpha;\nu \mu}
=\frac{v^{2}}{4}
\begin{pmatrix}
- \eta^{\nu\mu} & 0 & 0\\
0 & - \eta^{\nu\mu} & 0\\
0 & 0 & - \eta^{\nu\mu} - 2\delta^{\mu}_{0}\delta^{\nu}_{0}
\end{pmatrix},
\end{align}
which are consistent with Eq.~\eqref{eq:Seff}\footnote{The index of $\alpha$ and $\beta$ represents the index of broken generators, which is different from the index of fields. $\alpha=1,2$, and $3'$ correspond to $\chi^{2}$, $\chi^{1}$ and $\psi^{2}$, respectively.}.

\section{Summary and discussion}
\label{sec:summary}
We have derived low-energy coefficients of the inverse of retarded Green functions for Nambu--Goldstone modes in open systems, Eqs.~\eqref{eq:invG}-\eqref{eq:C10formula}, which provide a generalization of  the Nambu--Goldstone theorem. 
As is the case in isolated systems, we have classified the NG modes into type-A and type-B modes by using the coefficient of the single time-derivative term, $\rho^{\beta\alpha}$. These modes are further classified into diffusion and propagation modes. The relation between broken symmetries and the numbers of these modes are summarized in Eqs.~\eqref{eq:Type-B}, \eqref{eq:Type-A}, and \eqref{eq:relation}.  In this paper, we have employed continuum models. It is straightforward to generalize our results to systems with discrete translational symmetry, where the momentum is replaced by the Bloch momentum.

We have not taken into account hydrodynamic modes, which are expected to appear when $Q_{R}^{\alpha}$ is conserved. The coupling between NG and hydrodynamic modes may change low-energy behaviors. The limit $k\to 0$ in Eqs.~\eqref{eq:Cmu} and \eqref{eq:C10formula2} to obtain Eqs.~\eqref{eq:invG}-\eqref{eq:C10formula} might not be well defined, although our formulae in Eqs.~\eqref{eq:Gk}-\eqref{eq:DjRjA} still give the correct results.
 In such a case, it is better to treat the hydrodynamic modes as the dynamical degrees of freedom.
 
It is also fascinating to consider a spontaneous breaking of spacetime symmetries. In particular, time-translational symmetry breaking will be interesting in the context of ``quantum time crystal''~\cite{Wilczek:2012jt}. In isolated systems, it is shown that the time-translational symmetry cannot be broken in the ground state~\cite{Watanabe:2014hea}. On the other hand, in open systems, time-translationalsymmetry breaking can occur. 
For example, in reaction-diffusion systems, time-translational symmetry breaking is known as  synchronization phenomena, and the dynamics of phase, which corresponds to that of the NG mode, is described by a nonlinear diffusion equation~\cite{kuramoto2012chemical}. A similar situation can occur in open quantum systems~\cite{Hayata:2018qgt,Hongo:2018ant}.

 In the clarification of NG modes, we assumed $G_{\pi}^{-1}(\omega,\bm{k})=G_{\pi}^{-1}(\omega,-\bm{k})$, which leads to $C^{\beta\alpha;i}$ vanishing.  
 It will be interesting to relax this condition.  If there is a steady-state current that is an order parameter,  $C^{\beta\alpha;i}$ will be nonvanishing. In this case,  a different type of NG mode, which cannot be classified in this paper, will appear. 
 
 Another direction is to consider spontaneous symmetry breaking of higher-form symmetries~\cite{Gaiotto:2014kfa}. 
Photons can be understood as the NG modes associated with the spontaneous breaking of $U(1)$ one-form symmetry~\cite{Gaiotto:2014kfa}. 
As in the ordinary symmetry breaking,  there also exists the type-B photon, whose dispersion is quadratic~\cite{Yamamoto:2015maz,Ozaki:2016vwu,Sogabe:2019gif}.
It will be interesting to see how such a mode changes in an open system.  We leave these issues for future work.

\section*{Acknowledgements}
This work is supported in part by Japan Society of Promotion of Science (JSPS) Grant-in-Aid for Scientific Research
(KAKENHI Grants No. 16K17716, 17H06462, and 18H01211), by RIKEN  iTHEMS,
by the Zhejiang Provincial Natural Science Foundation Key Project (Grant No. LZ19A050001), and by NSF of China (Grant No. 11674283).

\bibliographystyle{ptephy}
\bibliography{ssb}

\appendix

\section{Proof of Eq.~\eqref{eq:classification}}\label{sec:NoTypeADiffusion}
We prove here that there is no type-A diffusion mode in the following type of  inverse of retarded Green function:
\begin{equation}
\begin{split}
[G_{\pi}^{-1}]^{\beta\alpha}(\omega,\bm{k})=- \ri \omega \rho^{\beta\alpha} -  \omega^{2}\bar{g}^{\beta\alpha}  + \bm{k}^{2}g^{\beta\alpha},
\label{eq:GpiInv}
\end{split}
\end{equation}
where  $\det g\neq0$ and $\det(-\ri\omega\rho-\omega^{2} \bar{g})\neq0$ at an arbitrary small $\omega$ are assumed.
At $\bm{k}=\bm{0}$, there are $(2N_{\text{BS}}-\rank \rho)$ solutions with $\omega=0$ to $\det G_{\pi}^{-1}(\omega,\bm{0})=0$.
As discussed in Sec. \ref{sec:Classification}, $2(N_\text{BS}-\rank \rho)$ and $\rank \rho$ solutions correspond to type-A and -B modes, respectively.
The type-B modes can be solved by neglecting the $\omega^{2}$ term in $G_{\pi}^{-1}(\omega,\bm{k})$,
i.e., $\det(- \ri \omega\rho + \bm{k}^{2} g)=0$. There are $\rank \rho$ solutions, which behaves like $\omega \sim |\bm{k}|^{2}$.

Let us next find the solutions for the type-A modes. For this purpose, it is useful to choose the basis of $G_{\pi}^{-1}$ such that 
\begin{equation}
\begin{split}
G_{\pi}^{-1}= \begin{pmatrix}
 - \omega^{2} \bar{g}_{1} + \bm{k}^{2}g_{1} &-\omega^{2}\bar{g}_{2}+\bm{k}^{2} g_{2}\\
-\omega^{2}\bar{g}_{3}+\bm{k}^{2}g_{3} & -\ri \omega\rho_{4} -\omega^{2} \bar{g}_{4}  +\bm{k}^{2}g_{4}
\end{pmatrix}
=: 
 \begin{pmatrix}
A & C\\
D & B
\end{pmatrix}.
\end{split}
\end{equation}
Roughly speaking, the submatrices $A$ and $B$ contain type-A and type-B modes; they interact through $C$ and $D$.
It is useful to employ the formula, $\det G_{\pi}^{-1}=( \det B )\det (A-CB^{-1}D)$ to find the dispersion relation for type A. 
If $CB^{-1}D$ is negligible, $\det A=0$ gives the solutions of type A. 
Since $A$ is an $(N_\text{BS}-\rank\rho)\times (N_\text{BS}-\rank\rho)$ matrix, 
the number of solutions is $2(N_\text{BS}-\rank\rho)$, as we expected.
Therefore, $\det A=0$ contain all type A NG modes. 
Since $\det A$ is a function of $\omega^{2}$, the solution has the form  $\omega = \pm a_{A}|\bm{k}|^{2}$. 
From the stability condition, $a_{A}$ must be real. Therefore, no diffusion modes appear in $\det A =0$. 
So far, we assumed that the contribution from $CB^{-1}D$ in $\det (A-CB^{-1}D)$ is negligible. Let us now check that it is correct.
When $\omega\sim |\bm{k}|$, the submatrices behave like $B\sim |\bm{k}|$, and $A\sim C\sim D\sim |\bm{k}|^{2}$. This leads to $CB^{-1}D\sim |\bm{k}|^{3}\ll  |\bm{k}|^{2}\sim A$. Therefore, the solutions are determined from $\det A=0$ at a small $\bm{k}$, and there is no type-A diffusion mode in Eq.~\eqref{eq:GpiInv}.

\end{document}